\documentclass[12pt]{article}
\usepackage{amsmath, amsthm, amssymb, bbm}
\usepackage{graphicx, float, subcaption}
\usepackage{booktabs, adjustbox, multirow, xcolor}
\usepackage{hyperref}
\usepackage{algorithm,algorithmic}
\usepackage[round]{natbib}

\newcommand{\blind}{0}

\theoremstyle{definition}
\newtheorem{definition}{Definition}

\begin{document}

\bibliographystyle{jasa3}

\def\spacingset#1{\renewcommand{\baselinestretch}%
{#1}\small\normalsize} \spacingset{1}

%%%%%%%%%%%%%%%%%%%%%%%%%%%%%%%%%%%%%%%%%%%%%%%%%%%%%%%%%%%%%%%%%%%%%%%%%%%%%%

\if0\blind
{
  \title{\bf Enhancing Sample Quality through \\ Minimum Energy Importance Weights}
  \author{
    Chaofan Huang \hspace{4mm} and \hspace{4mm} V. Roshan Joseph\thanks{Corresponding author: roshan@gatech.edu} \vspace{3mm} \\
    H. Milton Stewart School of Industrial and Systems Engineering, \\
    Georgia Institute of Technology, Atlanta, GA, 30332}
  \maketitle
} \fi

\if1\blind
{
  \bigskip
  \bigskip
  \bigskip
  \begin{center}
    {\LARGE\bf Title}
\end{center}
  \medskip
} \fi

\bigskip
\begin{abstract}
Importance sampling is a powerful tool for correcting the distributional mismatch in many statistical and machine learning problems, but in practice its performance is limited by the usage of simple proposals whose importance weights can be computed analytically. To address this limitation, \citet{liu2017bbis} proposed a Black-Box Importance Sampling (BBIS) algorithm that computes the importance weights for arbitrary simulated samples by minimizing the kernelized Stein discrepancy. However, this requires knowing the score function of the target distribution, which is not easy to compute for many Bayesian problems. Hence, in this paper we propose another novel BBIS algorithm using minimum energy design, BBIS-MED, that requires only the unnormalized density function, which can be utilized as a post-processing step to improve the quality of Markov Chain Monte Carlo samples. We demonstrate the effectiveness and wide applicability of our proposed BBIS-MED algorithm on extensive simulations and a real-world Bayesian model calibration problem where the score function cannot be derived analytically.
\end{abstract}

\noindent%
\vfill

\newpage
\spacingset{1.5} % DON'T change the spacing!

\section{Introduction}
\label{sec:introduction}

Bayesian inference is a popular tool for solving many statistical and engineering problems. However, in general, the closed-form derivation of the posterior $\pi$ is typically unattainable in practice, and thus approximation by Monte Carlo methods are often employed. Markov chain Monte Carlo (MCMC) and Importance Sampling (IS) are two popular techniques, but each with its own weakness. 

The key idea of MCMC \citep{metropolis1953mcmc,hastings1970mcmc} is to carefully construct a Markov chain that has the target posterior $\pi$ as the equilibrium distribution. Though easy to implement in practice, its asymptotic convergence is difficult to analyze and no finite-sample guarantee is known. To offset the random initialization and the auto-correlation within the MCMC samples, in practice it is recommended to discard the burn-in samples followed by thinning to yield the final sample set for the posterior approximation \citep{robert2013mc}. Hence, at least tens of thousands Markov chain iterations are required for obtaining sufficient samples, but this becomes a major computational bottleneck when the posterior $\pi$ is expensive to evaluate, e.g. the big data setting \citep{bardenet2017mcmc}, the calibration of complex computer code \citep{joseph2019mined}, etc., in which it would take days or even weeks to finish running the MCMC. 

On the other hand IS \citep{robert2013mc} instead draw samples from a simple proposal distribution $q$ and correct for the distributional mismatch by assigning the importance weight that is proportional to $\pi(\cdot)/q(\cdot)$. Improved convergence can be achieved if one can draw Quasi-Monte Carlo samples \citep{niederreiter1992qmc} from the proposal $q$ \citep{aistleitner2015qmc}. However, the performance of IS is sensitive to the choice of the proposal because a poor choice could possibly lead to an estimator with infinite variance \citep{owen2013mc}. It is challenging to design a good proposal $q$ for the high-dimensional complex posterior given the limited class of proposals that are analytically tractable. 

It is natural to consider combining the advantages of MCMC and IS: applying IS to correct for the distributional bias of the finite MCMC samples, which is believed to match more closely to the target posterior $\pi$ than the samples from the \emph{simple} proposal typically used by IS. This idea is explored in Markov Chain Importance Sampling \citep{rudolf2020mcis,schuster2021mcis}, which also allows for recycling the burn-in and thinned samples. However, it cannot be extended for the adaptive variant of MCMC that enjoys faster mixing in practice \citep{andrieu2008mcmc,vihola2012mcmc}. On the other hand, \citet{liu2017bbis} proposed the Black-Box Importance Sampling (BBIS) that can calculate the importance weights for samples simulated from arbitrary unknown black-box mechanisms. The importance weights are computed by minimizing over the kernelized Stein discrepancy \citep{liu2016ksd,chwialkowski2016ksd,oates2017ksd}, a goodness-of-fit measure for testing whether the samples are generated from the target distribution. However, computing the kernelized Stein discrepancy requires knowing the score function, i.e., the gradient of the log-density function $\nabla_{x}\log\pi(x)$. 

Unfortunately, the score function is not always available in many engineering applications, e.g., calibration of \emph{black-box} computer codes \citep{ngo1967fem,mak2018code}, where only input/output pairs are accessible. A much more common situation is where the score function exists but computing it is not straightforward, e.g., calibration for system of differential equations (Section~\ref{sec:bayesian_model_calibration}). See \citet{sengupta2014gcdm} for the existing numerical approaches on the gradient computation within the dynamical models. The adjoint method is suggested to be the computational and memory efficient approach, which involves solving another system of differential equations known as the sensitivity equation. Consequently, the expense of computing the gradient is at least on par with the cost of solving the system. This would certainly be a computational bottleneck for large-scale system, not to mention the potential numerical instability from solving the sensitivity equation. 

To address this shortcoming of kernelized Stein discrepancy-based BBIS algorithm, we propose BBIS-MED, a novel Black-Box Importance Sampling method using minimum energy design (MED) \citep{joseph2015mined} that only requires the unnormalized (log) density function, making it applicable for almost all Bayesian inference problems. BBIS-MED computes the importance weights by minimizing over an energy criterion, which has demonstrated its success as a good criterion for finding representative samples of any distribution \citep{joseph2019mined}. Moreover, extensive simulations are provided to demonstrate that our proposed BBIS-MED is orders of magnitude faster than the state-of-the-art BBIS algorithm, e.g., for a 55 dimensional logistic regression example in Subsection~\ref{subsec:logistic_regression_example}, BBIS-MED only takes 6 seconds to compute the importance weights for 5{,}000 samples, in contrast to the 1.3 hours required by the BBIS algorithm of \citet{liu2017bbis}.
 
The paper is organized as follows. Section~\ref{sec:minimum_energy_designs} reviews the minimum energy designs \citep{joseph2015mined,joseph2019mined} that motivates the BBIS-MED algorithm. Section~\ref{sec:minimum_energy_importance_weights} proposes the computational efficient BBIS-MED algorithm that only requires the unnormalized density function. Section~\ref{sec:exiting_black-box_importance_weights} outlines some existing BBIS algorithms from the literature, and extensive numerical comparisons to our proposed BBIS-MED is provided in Section~\ref{sec:simulations}. Section~\ref{sec:bayesian_model_calibration} presents a successful application of BBIS-MED on a real world Bayesian model calibration problem where the score function is not easy to compute. We conclude the article with some remarks in Section~\ref{sec:conclusion}.

\section{Minimum Energy Designs}
\label{sec:minimum_energy_designs}

Let us start by defining the minimum energy designs (MED) in \citet{joseph2015mined,joseph2019mined}, which forms the foundation of our proposed Black-Box Importance Sampling algorithm.

\begin{definition}[MED]
Let $\pi$ be any target density function over the support $\mathcal{X}\subseteq \mathbb{R}^{p}$. The $n$-point minimum energy design $\mathcal{D}^{*}_{n}$ of $\pi$ is defined as the optimal solution of 
\begin{align}
  \label{eq:med}
  \arg\min_{\mathcal{D}_n\in\mathbb{D}_n}E_{k}(\mathcal{D}_n) = \sum_{\substack{x_i,x_j\in\mathcal{D}_n\\i\neq j}}\left\{\frac{\rho(x_i)\rho(x_j)}{d(x_i,x_j)}\right\}^{k},
\end{align}
where $k > 0$, $\mathbb{D}_{n}=\{\{x_{i}\in\mathcal{X}\}_{i=1}^{n}\}$ is the set of all possible $n$-tuple over the support $\mathcal{X}$, $\rho(x) = 1/\pi^{1/(2p)}(x)$ is the charge function, and $d(x_{i},x_{j})$ is the Euclidean distance. 
\end{definition}

MED is motivated from the optimal configuration of charged particles in physics. It considers the design points $\mathcal{D}_{n}$ as some positively charged particles and aims to place them in the proper position for minimizing the total potential energy. Under the charge function $\rho(x) = 1/\pi^{1/(2p)}(x)$, \citet{joseph2019mined} show that for $k\to\infty$, the MED converges asymptotically to the target distribution $\pi$ as $n\to\infty$. 

MED is also known as the minimal Riesz energy points studied in the sphere-packing literature \citep{borodachov2008riesz1,borodachov2008riesz2,borodachov2019energy} by recognizing that $d(x_{i},x_{j})^{-1}$ is the Riesz $1$-kernel. Again \citet{borodachov2019energy} shows that using the charge function $\rho(x) = 1/\pi^{1/(2p)}(x)$ yields the optimal configuration that has the target distribution $\pi$ as the limiting distribution, while only requiring $k>p$. Similar convergence result is also presented in \citet{wang2011med}. Given that the optimal configuration that minimizes \eqref{eq:med} converges to $\pi$ as $n\to\infty$ when the charge function $\rho(x) = 1/\pi^{1/(2p)}(x)$, in the limit we should have 
\begin{align}
  \label{eq:cmed}
  \pi = \arg\min_{\mu}\mathbb{E}_{x_{i},x_{j}\sim\mu}\left[\left\{\frac{\rho(x_i)\rho(x_j)}{d(x_i,x_j)}\right\}^{k}\right],
\end{align}
where $\mu$ is any distribution supported on $\mathcal{X}$. The MED formulation \eqref{eq:med} can be generalized to arbitrary symmetric proper kernel $K:\mathcal{X}\times\mathcal{X}\to\mathbb{R}\cup\{+\infty\}$, but less is known about the limiting distribution of the optimal configuration beyond the Riesz kernel \citep{borodachov2019energy}.

\section{Minimum Energy Importance Weights}
\label{sec:minimum_energy_importance_weights}

Let $\{x_{i}\}_{i=1}^{n}$ be a set of samples in $\mathbb{R}^{p}$ that is simulated from an arbitrary (unknown) mechanism. The Black-Box Importance Sampling (BBIS) algorithm aims to find the normalized weights $\{w_{i}\}_{i=1}^{n}$, i.e., $w_{i}\geq 0$ and $\sum_{i=1}^{n}w_{i}=1$, such that the weighted samples $\{(x_{i},w_{i})\}_{i=1}^{n}$ give a good approximation to the target distribution $\pi$. Unlike \citet{liu2017bbis} that use kernelized Stein discrepancy to measure the similarity between $\{(x_{i},w_{i})\}_{i=1}^{n}$ and $\pi$, we instead consider the energy criterion in \eqref{eq:cmed} that  require knowing $\pi$ only up to a constant of proportionality, making it applicable to almost all Bayesian problems. 

More specifically, the Monte Carlo approximation for the expectation in \eqref{eq:cmed}, the continuous energy criterion, using weighted samples $\{(x_{i},w_{i})\}_{i=1}^{n}$ is
\begin{align}
  \label{eq:wmed1}
  \sum_{i=1}^{n}\sum_{j=1}^{n}w_{i}w_{j}\left\{\frac{\rho(x_i)\rho(x_j)}{d(x_i,x_j)}\right\}^{k}.
\end{align}
Compared to the MED formulation in \eqref{eq:med}, the approximation using weighted samples in \eqref{eq:wmed1} includes the self-interaction terms, i.e., terms with $i=j$, in the summation. Without these self-interaction terms, the minimum value of \eqref{eq:wmed1} is always obtained at 0 with weights $w_{i}=1$ for some $i$ and $w_{j}=0$ for all $j\neq i$. This is expected since if there is only one particle in the system, there will be no interaction (repulsion), and the total potential energy will always be minimized at 0. Hence, the self-interaction terms are required when weighted samples are used for approximating the energy criterion. Moreover, the self-interaction terms can also serve as a penalization to avoid the weights collapsing to only a few samples. 

However, after including the self-interaction terms, \eqref{eq:wmed1} always yields infinity for any normalized weights $\{w_{i}\}_{i=1}^{n}$ since $d(x_{i},x_{j}) = 0$ for $i=j$. One fix is to add some small positive term $\delta$ to the distance computation to bound it away from 0, i.e, we consider the following formulation, 
\begin{align}
  \label{eq:wmed2}
  \sum_{i=1}^{n}\sum_{j=1}^{n}w_{i}w_{j}\left\{\frac{\rho(x_i)\rho(x_j)}{(d^{2}(x_i,x_j)+\delta)^{1/2}}\right\}^{k},
\end{align}
where the Riesz kernel in \eqref{eq:wmed1} is replaced by the inverse multiquadric kernel $K(x_{i},x_{j}) = (d^{2}(x_i,x_j)+\delta)^{-1/2}$ for some small $\delta > 0$. Unfortunately, there is no known result on the limiting distribution for the optimal configuration of MED for the inverse multiquadric kernel. However, given that the inverse multiquadric kernel with small $\delta$ closely mimics the Riesz kernel except when $d(x_{i},x_{j})\approx 0$, one can assume that the difference between the limiting distribution of the optimal configuration with respect to the inverse multiquadric kernel and the limiting distribution with respect to the Riesz kernel should be very small. Hence, utilizing the limiting behavior of MED with respect to the Riesz kernel discussed in Section~\ref{sec:minimum_energy_designs}, one can believe that for $\rho(x)=1/\pi^{1/(2p)}(x)$, the minimal value of \eqref{eq:wmed2} with respect to the weights $\{w_{i}\}_{i=1}^{n}$ is obtained when the weighted samples $\{(x_{i},w_{i})\}_{i=1}^{n}$ are approximately following $\pi$. Recall that $\{w_{i}\}_{i=1}^{n}$ serve as the self-normalized importance weights that correct for the mismatch between the target distribution $\pi$ and the unknown underlying distribution of $\{x_{i}\}_{i=1}^{n}$.

This leads to the following optimization problem in matrix form for computing the black-box importance weights $w^{*}=(w^{*}_{1},\ldots,w^{*}_{n})$ that matches any samples $\{x_{i}\}_{i=1}^{n}$ to the target distribution $\pi$ via minimizing the energy criterion, 
\begin{align}
  \label{eq:meis}
  w^{*} = \arg\min_{w\in\Delta^{n}} w^{T}Rw,
\end{align}
where $\Delta^{n}$ is the $n$-dimensional probability simplex and
\begin{align*}
  R_{ij} =& \left\{\frac{\rho(x_i)\rho(x_j)}{(d(x_i,x_j)^2+\delta)^{1/2}}\right\}^{k} \\ 
  =& \exp\bigg\{-k\bigg(\frac{1}{2p}\log\pi(x_{i})+\frac{1}{2p}\log\pi(x_{j}) + \frac{1}{2}\log(d^{2}(x_i,x_j)+\delta)\bigg)\bigg\}, \\
  =& C\exp\bigg\{-k\bigg(\frac{1}{2p}\log\gamma(x_{i})+\frac{1}{2p}\log\gamma(x_{j}) + \frac{1}{2}\log(d^{2}(x_i,x_j)+\delta)\bigg)\bigg\},
\end{align*}
where $\gamma(x) = \pi(x)/Z$ is the unnormalized density function and $C = Z^{k/p}$ is some constant depending on the unknown normalizing constant $Z$. Since the constant $C$ can be factored out without changing the optimum solution for \eqref{eq:meis}, our proposed BBIS algorithm only requires the \emph{unnormalized log-density function}, which is the most elementary prerequisite for any sampling method. Moreover, the optimization problem in \eqref{eq:meis} is a convex quadratic programming problem (see Appendix \ref{appendix:pd_of_R}) that can be solved efficiently with theoretical convergence guarantee by the simplex constrained convex optimization techniques, including mirror descent \citep{nemirovskij1983md,beck2003md}, projected gradient descent \citep{duchi2008projgrad}, and Frank-Wolfe \citep{jaggi2013frank-wolfe}. We employ projected gradient descent for the simulations presented in this paper. 

Last, let us discuss the choices for the parameters in \eqref{eq:meis}. 
\begin{itemize}
  \item For the distance function $d(x_{i},x_{j})$, we use the Mahalanobis distance, 
  \begin{align*}
    d^{2}(x_{i},x_{j}) = (x_{i} - x_{j})^{T}\Sigma^{-1}(x_{i}-x_{j}),
  \end{align*}
  where $\Sigma$ is the covariance matrix estimated from the samples $\{x_{i}\}_{i=1}^{n}$. It has been shown in \citep{joseph2019mined} that for computing MED of correlated distributions, better empirical performance is observed from Mahalanobis distance compared to Euclidean distance. 
  \item For $\delta$, we find that after standardization of the samples $\{x_{i}\}_{i=1}^{n}$ to be component-wise zero mean and unit variance, using $\delta = 0.01$ yields robust empirical performance on problems across different dimensions. 
  \item For $k$, the theoretical result of MED from Section~\ref{sec:minimum_energy_designs} suggests that one should use $k > p$, but empirically the performance is not good (see Appendix \ref{appendix:choice_for_k}). Our guess is that the theoretical results no longer hold when the self-interaction terms are included in the energy criterion. Using $k=1$ gives the most robust performance from extensive numerical simulations.
\end{itemize}

Given that our proposed BBIS algorithm is motivated from MED, for the rest of paper we will refer it as BBIS-MED and the resulting weights $w^{*}$ as the minimum energy importance weights. Though the theoretical guarantee of BBIS-MED requires future study, which is known to be a very difficult mathematical problem, extensive simulations are provided in the later Sections to demonstrate that any arbitrary simulated samples $\{x_{i}\}_{i=1}^{n}$ with the minimum energy importance weights $w^{*}$ provide better approximation for the target distribution $\pi$ than the unweighted samples $\{x_{i}\}_{i=1}^{n}$ alone. 

\section{Existing Black-Box Importance Weights}
\label{sec:exiting_black-box_importance_weights}

Let us now review some existing BBIS algorithms in the literature that we will compare BBIS-MED to empirically in Section~\ref{sec:simulations}. 

\subsection{Stein Importance Weights}
\label{subsec:stein_importance_weights}

The Stein Importance weights \citep{liu2017bbis,hodgkinson2020ksd} is built upon the kernelized Stein discrepancy \citep{liu2016ksd,chwialkowski2016ksd,oates2017ksd}, an integral probability metric (IPM) for measuring how well the samples $\{x_{i}\}_{i=1}^{n}$ cover the domain $\mathcal{X}$ with respect to some target distribution $\pi$. Let $k:\mathcal{X}\times\mathcal{X}\to\mathbb{R}$ be a reproducing kernel of reproducing kernel Hilbert spaces \citep{hickernell1998rkhs} of functions $\mathcal{X}\to\mathbb{R}$. The kernelized Stein discrepancy (KSD) for distribution $\pi$ and $q$ over the same support $\mathcal{X}$ is defined as 
\begin{align}
  \label{eq:ksd}
  \mathbb{S}(q,\pi) = \mathbb{E}_{x,x^{\prime}\sim q}[k_{\pi}(x,x^{\prime})],
\end{align}
where $k_{\pi}(x,x^{\prime})$ is a kernel function defined by  
\begin{align*}
  k_{\pi}(x,x^{\prime}) =& s_{\pi}(x)^{T}k(x,x^{\prime})s_{\pi}(x^{\prime}) + s_{\pi}(x)^{T}\nabla_{x^{\prime}}k(x,x^{\prime}) + \\
  & s_{\pi}(x^{\prime})^{T}\nabla_{x}k(x,x^{\prime}) + \text{trace}(\nabla_{x,x^{\prime}}k(x,x^{\prime})),
\end{align*}
where $s_{\pi}(x)=\nabla_{x} \log \pi(x)$ is the score function of $\pi(x)$. \citet{liu2016ksd} shows that $\mathbb{S}(\pi,q)$ is always non-negative since $k_{\pi}(x,x^{\prime})$ is positive definite if $k(x,x^{\prime})$ is positive definite. Moreover, $\mathbb{S}(\pi,q)=0$ if and only if $\pi=q$ under some mild conditions on the kernel $k(x,x^{\prime})$ that are satisfied by most commonly used kernels such as the radial basis function (RBF) and inverse multiquadric kernel. 

It follows that the empirical version of KSD in \eqref{eq:ksd} for measuring the discrepancy between the weighted samples $\{(x_{i},w_{i})\}_{i=1}^{n}$ and the target distribution $\pi$ is
\begin{align}
  \label{eq:wksd}
  \mathbb{S}(\{(x_{i},w_{i})\}_{i=1}^{n},\pi) = \sum_{i=1}^{n}\sum_{j=1}^{n}w_{i}w_{j}k_{\pi}(x_{i},x_{j}).
\end{align}
Hence, the optimal weights $w^{*}=(w^{*}_{1},\ldots,w^{*}_{n})$ such that $\{(x_{i},w^{*}_{i})\}_{i=1}^{n}$ best approximate $\pi$ would be the weights that minimize \eqref{eq:wksd}, leading to following optimization problem proposed in \citet{liu2017bbis} to compute $w^{*}$, 
\begin{align}
  \label{eq:bbis-ksd}
  w^{*} = \arg\min_{w\in\Delta^{n}} w^{T}K_{\pi}w,
\end{align}
where $K_{\pi} = \{k_{\pi}(x_{i},x_{j})\}_{i,j=1}^{n}$ and $w=(w_{1},\ldots,w_{n})$. Note that \eqref{eq:bbis-ksd} is also a convex quadratic programming problem, which can be solved efficiently by the existing optimization tools. We refer this algorithm as BBIS-KSD for the rest of the paper. For implementation, we use the R package \texttt{stein.thinning} \citep{chen2021stein} to compute $K_{\pi}$ in \eqref{eq:bbis-ksd}. Inverse multiquadric kernel $k(x,x^{\prime}) = (1+\lVert \Gamma^{-1/2}(x-y)\rVert^{2})^{-1/2}$ is used where $\Gamma=\ell^2 I$ with $\ell$ chosen by the median heuristic \citep{gretton2012mmd,garreau2017mh}, the median of the pairwise Euclidean distance between samples $\{x_{i}\}_{i=1}^{n}$. 

There is also related work by \citet{oates2017ksd} applying the Steinalized kernel $k^{+}_{\pi}(x,x^{\prime}) = k_{\pi}(x,x^{\prime}) + 1$ in control variate to compute the importance weights, which is equivalent to Bayesian Monte Carlo \citep{rasmussen2003bmc} with kernel $k^{+}_{\pi}(x,x^{\prime})$. Since it also belongs to the class of BBIS algorithms that require knowing the score function of $\pi$ and its empirical performance is comparable to BBIS-KSD as shown in \citet{liu2017bbis}, we only provide the numerical comparison to BBIS-KSD in the simulation studies. 

Although the theoretical understanding of KSD is rich \citep{liu2016ksd,chwialkowski2016ksd,oates2017ksd,liu2017bbis}, computing KSD requires knowing the score function of $\pi$, which limits its applicability for many Bayesian problems, e.g., calibration of complex computer codes, where we cannot analytically derive the score function. 

\subsection{KDE Importance Weights}
\label{subsec:kde_importance_weights}

Another approach for computing the importance weights $\{w^{*}_{i}\}_{i=1}^{n}$ is to first construct an estimator $\hat{q}$ for the unknown proposal $q$ that simulated the samples $\{x_{i}\}_{i=1}^{n}$, and then compute the normalized weights using $\hat{q}$, i.e.,
\begin{align*}
  w^{*}_{i} = \frac{\pi(x_{i})/\hat{q}(x_{i})}{\sum_{j=1}^{n}\pi(x_{j})/\hat{q}(x_{j})}.
\end{align*}
\citet{henmi2007bbis} consider using parametric family for $\hat{q}$ with the parameters obtained by maximum likelihood estimation. \citet{delyon2016bbis} later extended this idea to non-parametric setting by using the leave-one-out kernel density estimator (KDE) for $\hat{q}$, i.e., 
\begin{align*}
  \hat{q}(x_{i}) = \frac{1}{n-1}\sum_{1\leq j \leq n, j\neq i}k(x_{i},x_{j}).
\end{align*}
Surprisingly, under certain smoothness assumptions, \citet{henmi2007bbis} and \citet{delyon2016bbis} show that using estimator $\hat{q}$ could lead to improved weights than the standard IS weights computing from the exact $q$.  

For the comparison to BBIS-MED, we focus on the KDE estimator studied in \citet{delyon2016bbis}. As suggested in \citet{liu2017bbis}, RBF kernel is used with the rule-of-thumb bandwidth, 
\begin{align*}
  h=\hat{\sigma}\left(\frac{p2^{p+5}\Gamma(p/2+3)}{(2p+1)n}\right)^{1/(4+p)},
\end{align*}
where $\hat{\sigma}$ is the standard deviation of $\{x_{i}\}_{i=1}^{n}$ and $\Gamma(\cdot)$ is the Gamma function. We refer this method as BBIS-KDE for simplicity. 

\section{Simulations}
\label{sec:simulations}

In this section we provide extensive simulations comparing our proposed BBIS-MED to the two other BBIS algorithms discussed in Section~\ref{sec:exiting_black-box_importance_weights}: BBIS-KSD and BBIS-KDE. 

\subsection{Multivariate Gaussian Example}
\label{subsec:multivariate_gaussian_example}

Consider multivariate Gaussian $\pi=\mathcal{N}(0,\Sigma)$ as the target distribution. We test the performance of the BBIS algorithms under different sample size $n$, different correlation parameter $\tau\in[0,1]$ of the covariance matrix $\Sigma_{ij} = \tau^{|i-j|}$, and different problem dimension $p$. 

\paragraph{Evaluation Metric} 
To measure how well the weighted samples $\{x_{i},w_{i}\}_{i=1}^{n}$ approximate the target distribution $\pi$, we consider the energy distance,  a statistical potential proposed by \citet{szekely2004ed} for testing goodness-of-fit in multi-dimension. The energy distance between $\{x_{i},w_{i}\}_{i=1}^{n}$ and $\pi$ is defined as 
\begin{equation}
  \begin{split}
    \label{eq:ed}
      \mathcal{E}(\{x_{i},w_{i}\}_{i=1}^{n}, \pi) =& 2\sum_{i=1}^{n}w_{i}\mathbb{E}_{Y\sim\pi}\lVert x_{i} - Y\rVert_{2} - \sum_{i=1}^{n}\sum_{j=1}^{n}w_{i}w_{j}\lVert x_{i} - x_{j}\rVert_{2} - \\
      & \mathbb{E}_{Y,Y^{\prime}\sim\pi} \lVert Y - Y^{\prime}\rVert_{2}.
  \end{split}
\end{equation}
Similar to the kernelized Stein discrepancy, energy distance is also non-negative and equal to 0 if and only if the two distributions are identical. Moreover, Theorem 4 of \citet{mak2018sp} shows that for a large class of integrand $\phi$, the squared integration error is upper bounded by a term proportional to the energy distance, i.e., 
\begin{align*}
  \left(\sum_{i=1}^{n}w_{i}\phi(x_{i}) - \mathbb{E}_{\pi}\phi(X)\right)^2 \leq C_{\phi}\mathcal{E}(\{x_{i},w_{i}\}_{i=1}^{n}, \pi),
\end{align*}
where $C_{\phi}$ is a constant depending only on the integrand $\phi$. This shows the connection of energy distance to the integration error, the key objective of Monte Carlo approximation, verifying that the energy distance is a suitable evaluation metric for our problem. However, the expectation in \eqref{eq:ed} cannot be usually computed in closed-form, and thus Monte Carlo approximation is used by simulating large samples $\{y_{m}\}_{m=1}^{M}\sim\pi$. We can also drop the last term $\mathbb{E}_{Y,Y^{\prime}\sim\pi} \lVert Y - Y^{\prime}\rVert_{2}$ of \eqref{eq:ed} since it is a constant with respect to the weights $\{w_{i}\}_{i=1}^{n}$, leading to the following simplified metric for performance evaluation, 
\begin{align}
  \label{eq:mced}
    \hat{\mathcal{E}}(\{x_{i},w_{i}\}_{i=1}^{n}, \pi) = \frac{2}{M}\sum_{i=1}^{n}\sum_{m=1}^{M} w_{i}\lVert x_{i} - y_{m}\rVert_{2} - \sum_{i=1}^{n}\sum_{j=1}^{n}w_{i}w_{j}\lVert x_{i} - x_{j}\rVert_{2}.
\end{align}
For the multivariate Gaussian $\pi = \mathcal{N}(0,\Sigma)$ example, we simulate $M=10^{6}$ inverse Sobol' points \citep{bratley1988sobol} for the large samples $\{y_{m}\}_{m=1}^{M}$. Energy distance is preferred here for a fair comparison to avoid using other goodness-of-fit measures that are already serving the basis for the BBIS algorithms.   

\paragraph{Simulation Mechanism}
Now we present the mechanism that generate samples $\{x_{i}\}_{i=1}^{n}$. As suggested in \citet{liu2017bbis}, we only need the empirical distribution of $\{x_{i}\}_{i=1}^{n}$ to be ``roughly" close to the target distribution $\pi$, and the black-box weights will help correct for the mismatch. Hence, in the simulation studies we run the adapted MCMC \citep{vihola2012mcmc} for $n$ iterations and use all \emph{proposal} samples (including the \emph{rejected} ones) for $\{x_{i}\}_{i=1}^{n}$. The MCMC chain is initialized at some random sample $x_{1}\sim\mathcal{N}(0,I_{p})$, and the initial MCMC kernel covariance is set to $I_{p}$. The reason we go with MCMC is to best mimic the applications in real world problems where we often cannot directly sample from the posterior distribution, i.e., via the inverse transformation method. 

\paragraph{Baseline} 
Given that the samples $\{x_{i}\}_{i=1}^{n}$ are simulated from MCMC, one good baseline comparison for the BBIS algorithms are the \emph{accepted} MCMC samples each with $1/n$ weight. We do not discard any burn-in samples since we purposely start the MCMC with good initialization. We refer this as the MCMC baseline in the comparisons.

\begin{figure}[!t]
\centering
\begin{subfigure}{\textwidth}
    \centering
    \includegraphics[width=0.48\linewidth]{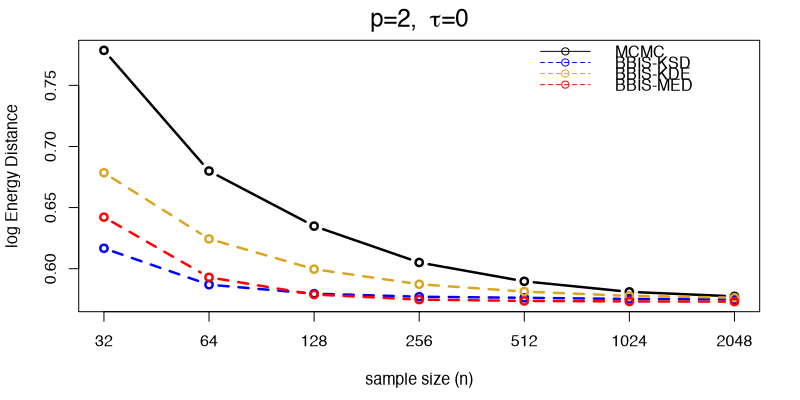}
    \hfill
    \includegraphics[width=0.48\linewidth]{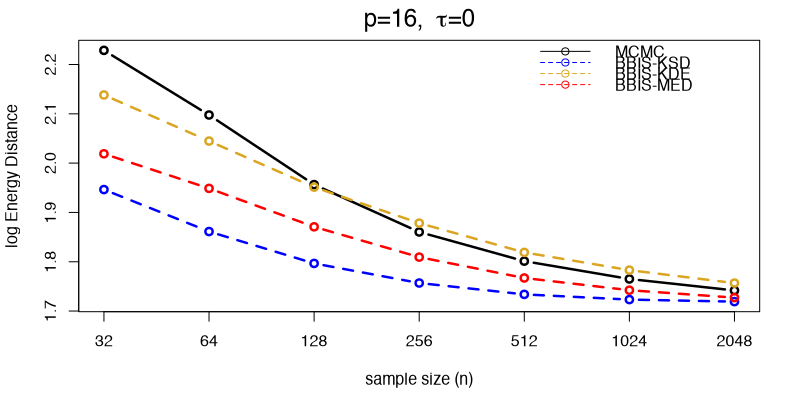}
    \caption{Varying by sample size $n$.}
    \label{fig:gaussian_edist_n}
\end{subfigure}

\begin{subfigure}{\textwidth}
    \centering
    \includegraphics[width=0.48\linewidth]{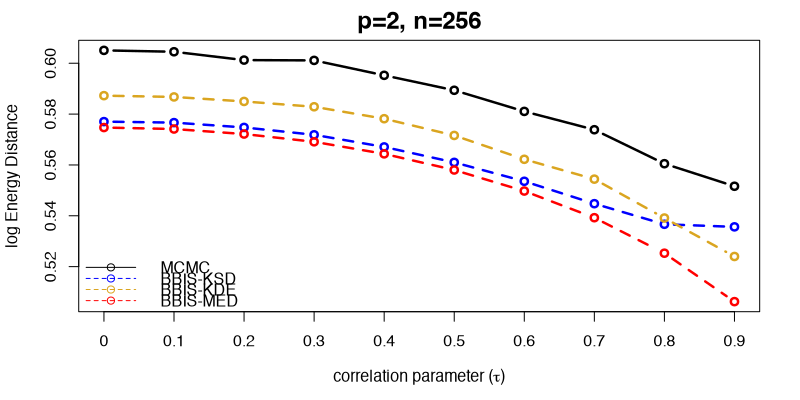}
    \hfill
    \includegraphics[width=0.48\linewidth]{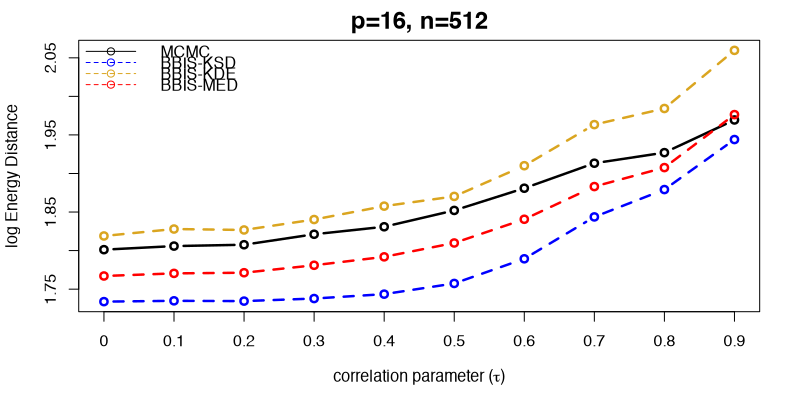}
    \caption{Varying by correlation parameter $\tau$.}
    \label{fig:gaussian_edist_tau}
\end{subfigure}

\begin{subfigure}{\textwidth}
    \centering
    \includegraphics[width=0.48\linewidth]{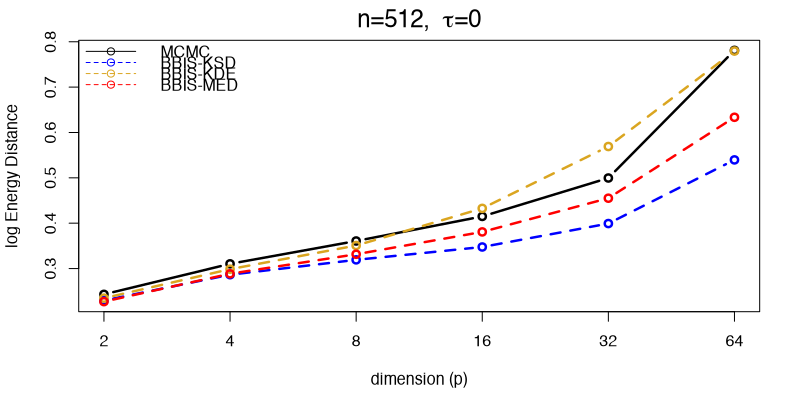}
    \hfill
    \includegraphics[width=0.48\linewidth]{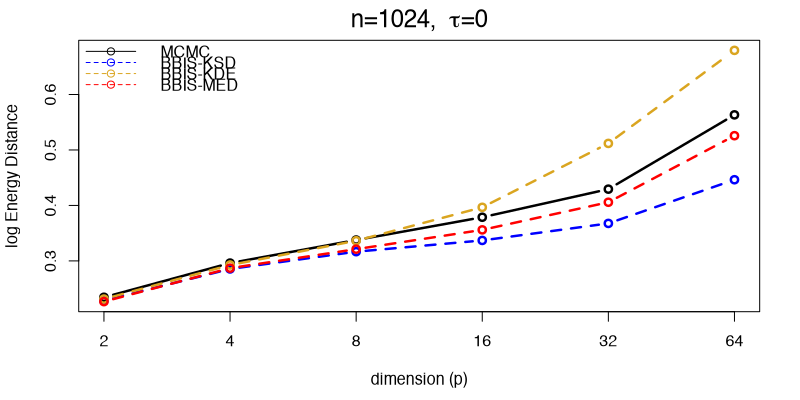}
    \caption{Varying by dimension $p$. Energy distance is normalized by dimension, i.e. divided by $\sqrt{p}$.}
    \label{fig:gaussian_edist_p}
\end{subfigure}

\caption{Comparison of the BBIS algorithms to the MCMC baseline on the multivariate Gaussian example with different (\subref{fig:gaussian_edist_n}) sample size $n$, (\subref{fig:gaussian_edist_tau}) correlation parameter $\tau$, and (\subref{fig:gaussian_edist_p}) dimension $p$. The energy distance \eqref{eq:mced} is averaged over 100 independent runs.}
\label{fig:gaussian_edist}
\end{figure}

\paragraph{Varying Sample Size}
Figure~\ref{fig:gaussian_edist_n} shows the comparison of the BBIS algorithms to the MCMC baseline for different sample size $n$. We can see that the BBIS-MED and BBIS-KSD outperform the baseline MCMC for all sample sizes on both small ($p=2$) and moderate ($p=16)$ dimensions. Nevertheless we also observe the expected diminishing improvement as the sample size $n$ increases. 

\paragraph{Varying Correlation Parameter}
Figure~\ref{fig:gaussian_edist_tau} shows the comparison of the BBIS algorithms to the MCMC baseline for different correlation parameter $\tau$. Again we also observe that both BBIS-MED and BBIS-KSD improve over the baseline MCMC samples, and stronger correlation does not impact the performance much, but the improvement is more significant on the smaller dimensional ($p=2$) problem. 

\paragraph{Varying Dimension}
Figure~\ref{fig:gaussian_edist_p} shows the comparison of the BBIS algorithms to the MCMC baseline for different dimension $p$ under fixed sample size $n$ and correlation parameter $\tau=0$. Similar to prior observations, BBIS-MED and BBIS-KSD yield weighted samples that better approximate the target distribution $\pi$ than the baseline MCMC samples. On higher dimensions, BBIS-KSD wins over our proposed BBIS-MED, which is expected since BBIS-KSD has access to more information, i.e., the score function of $\pi$.

\begin{figure}[!t]
\centering
\begin{subfigure}{0.48\textwidth}
    \centering
    \includegraphics[width=\textwidth]{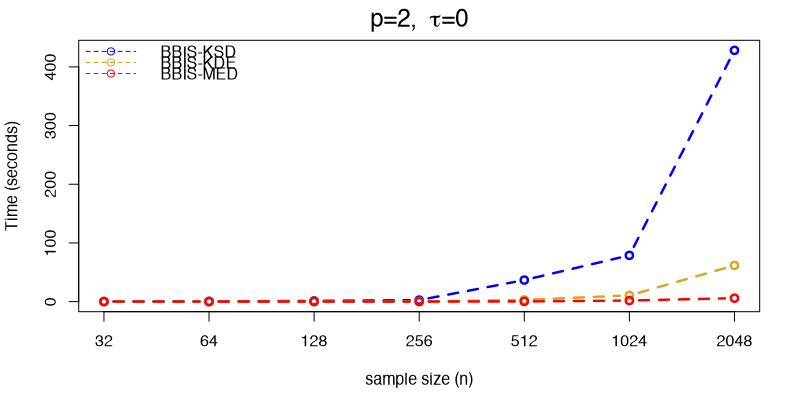}
    \caption{Varying by sample size $n$.}
    \label{fig:gaussian_times_n}
\end{subfigure}
\begin{subfigure}{0.48\textwidth}
    \centering
    \includegraphics[width=\linewidth]{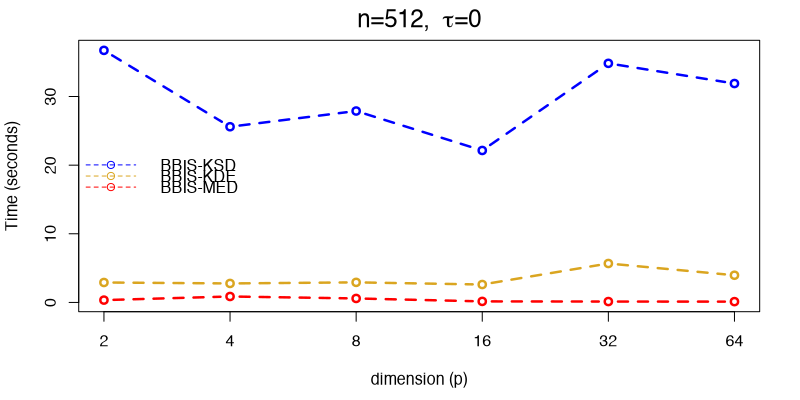}
    \caption{Varying by dimension $p$.}
    \label{fig:gaussian_times_p}
\end{subfigure}
\caption{Computational time of the BBIS algorithms on the multivariate Gaussian example averaged over 100 independent runs.}
\label{fig:gaussian_times}
\end{figure}

\paragraph{Computational Time}
Figure~\ref{fig:gaussian_times} compares the computational efficiency of the BBIS algorithms. Our proposed BBIS-MED is the clear winner this time and it is order of magnitude faster than the BBIS-KSD, e.g. for sample size $n=2048$, BBIS-MED takes only around 6 seconds whereas BBIS-KSD needs 428 seconds. The computational complexity of BBIS-MED is quadratic in sample size $n$ and linear in the dimension $p$ from computing the pairwise distance, but interestingly from Figure~\ref{fig:gaussian_times}, the actual runtime is almost constant as dimension increases (right panel), owing to the fast pairwise distance computation in R.

\paragraph{Summary} BBIS-MED and BBIS-KSD both beat the MCMC baseline on the multivariate Gaussian example, while BBIS-KDE could sometimes lead to worse approximation for the target distribution $\pi$. Though the performance of BBIS-MED is not as good as BBIS-KSD in higher dimensions, BBIS-MED (i) enjoys significant computational saving and (ii) is applicable to a much broader class of Bayesian problems where we do not require computing the score function of $\pi$. Additional simulation results on other evaluation metrics such as the integral approximation errors on a few test functions are provided in Appendix~\ref{appendix:additional_simulation_multivariate_gaussian}, that is to assess the approximation accuracy of $\mathbb{E}_{X\sim\pi}[\phi(X)]$ for some integrand $\phi$ by $\sum_{i=1}^{n}w_{i}\phi(x_{i})$, where the samples $\{x_{i}\}_{i=1}^{n}$ are generated using the adapted MCMC and the weights $\{w_{i}\}_{i=1}^{n}$ are computed by the BBIS algorithms. 

\begin{figure}[!t]
\centering
\begin{subfigure}{\textwidth}
    \centering
    \includegraphics[width=0.48\linewidth]{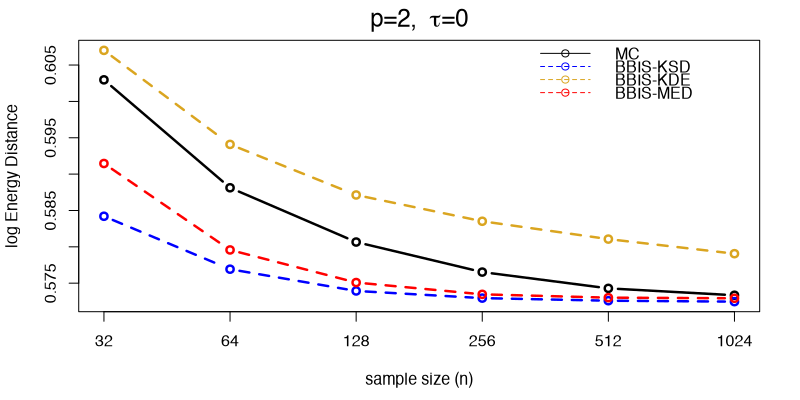}
    \hfill
    \includegraphics[width=0.48\linewidth]{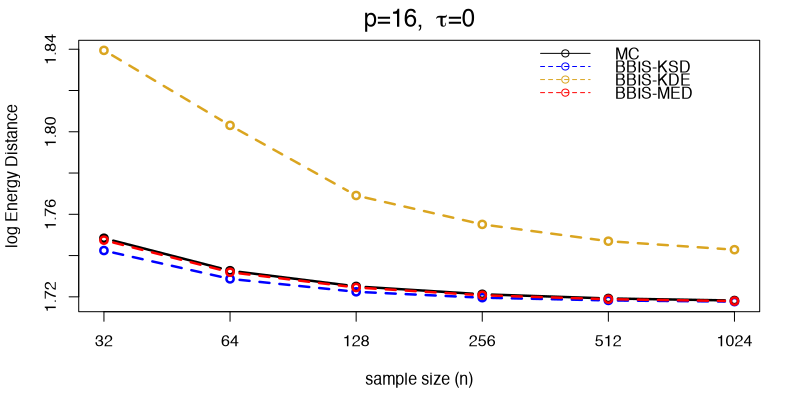}
    \caption{Varying by sample size $n$.}
    \label{fig:gaussian_mc_edist_n}
\end{subfigure}

\begin{subfigure}{\textwidth}
    \centering
    \includegraphics[width=0.48\linewidth]{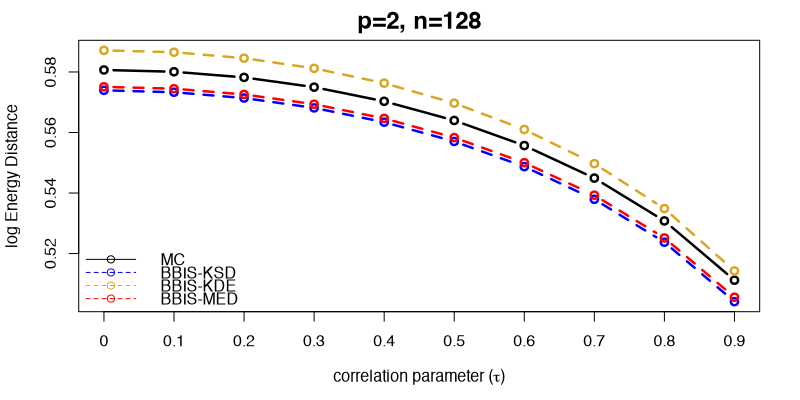}
    \hfill
    \includegraphics[width=0.48\linewidth]{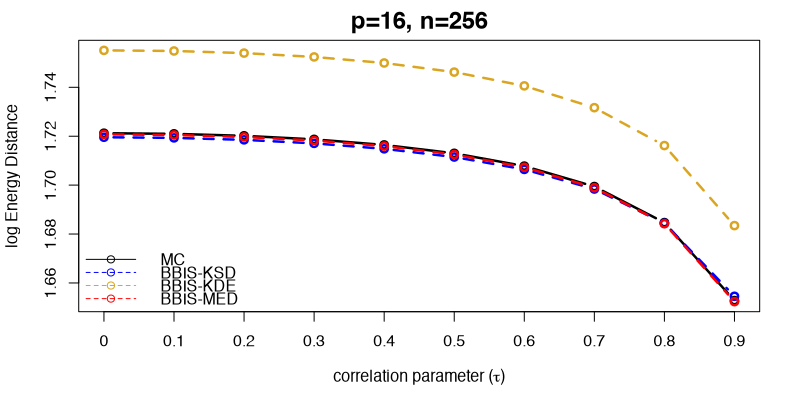}
    \caption{Varying by correlation parameter $\tau$.}
    \label{fig:gaussian_mc_edist_tau}
\end{subfigure}

\begin{subfigure}{\textwidth}
    \centering
    \includegraphics[width=0.48\linewidth]{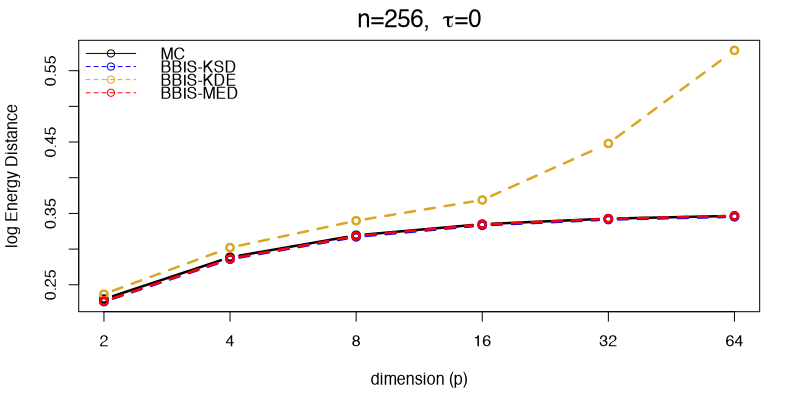}
    \hfill
    \includegraphics[width=0.48\linewidth]{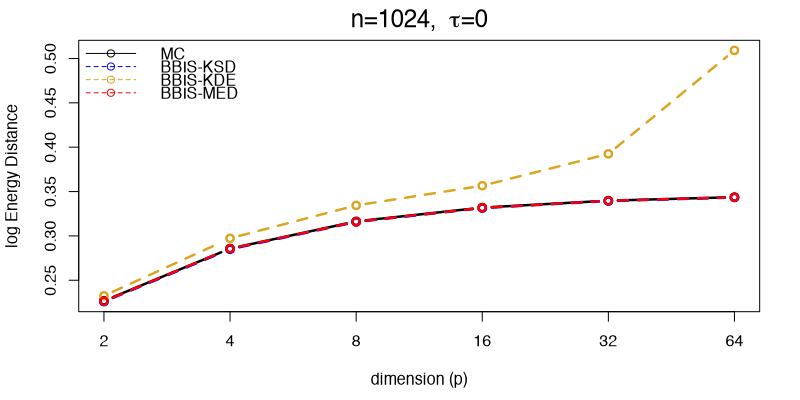}
    \caption{Varying by dimension $p$. Energy distance is normalized by dimension, i.e. divided by $\sqrt{p}$.}
    \label{fig:gaussian_mc_edist_p}
\end{subfigure}

\caption{Comparison of the BBIS algorithms to the MC baseline on the multivariate Gaussian example with different (\subref{fig:gaussian_mc_edist_n}) sample size $n$, (\subref{fig:gaussian_mc_edist_tau}) correlation parameter $\tau$, and (\subref{fig:gaussian_mc_edist_p}) dimension $p$. The energy distance \eqref{eq:mced} is averaged over 100 independent runs.}
\label{fig:gaussian_mc_edist}
\end{figure}

\paragraph{Monte Carlo Baseline} In addition, given that direct sampling is feasible for the multivariate Gaussian distribution, it would be interesting to assess the performance of the BBIS algorithms on the Monte Carlo (MC) samples of the multivariate Gaussian distribution. The unweighted Monte Carlo samples $\{x_{i}\}_{i=1}^{n}$ serve as the baseline for this comparison, and we refer it as the MC baseline. The energy distance \eqref{eq:mced} is again used as the evaluation metric. Similar findings are observed from Figure~\ref{fig:gaussian_mc_edist}: both BBIS-MED and BBIS-KSD outperform the MC baseline, albeit the improvement diminishes for higher dimensional problems. On the other hand, given that we are not able to provide the theoretical guarantee of the BBIS-MED at this time, its comparison to the MC baseline demonstrates that BBIS-MED could still improve (or at least not deteriorate) the approximation quality, which serves as a good empirical validation for the correctness of our proposed method.
 
\subsection{Mixture Gaussian Example}
\label{subsec:mixture_gaussian_example}

\begin{figure}[!t]
\centering
\begin{subfigure}{0.48\textwidth}
    \centering
    \includegraphics[width=\textwidth]{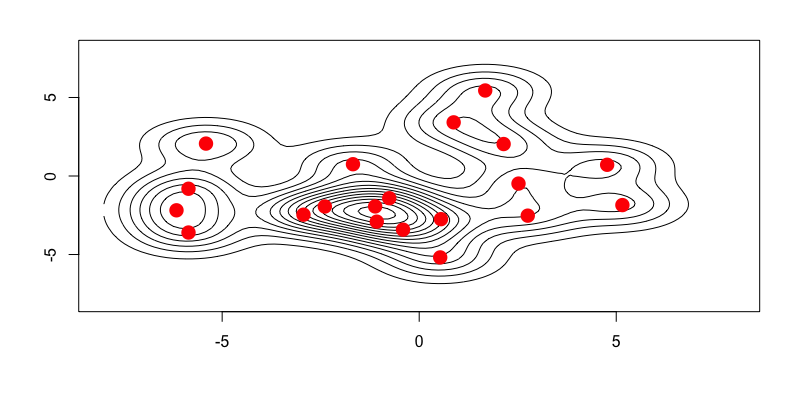}
    \caption{Contour of $\pi$.}
\end{subfigure}
\begin{subfigure}{0.48\textwidth}
    \centering
    \includegraphics[width=\textwidth]{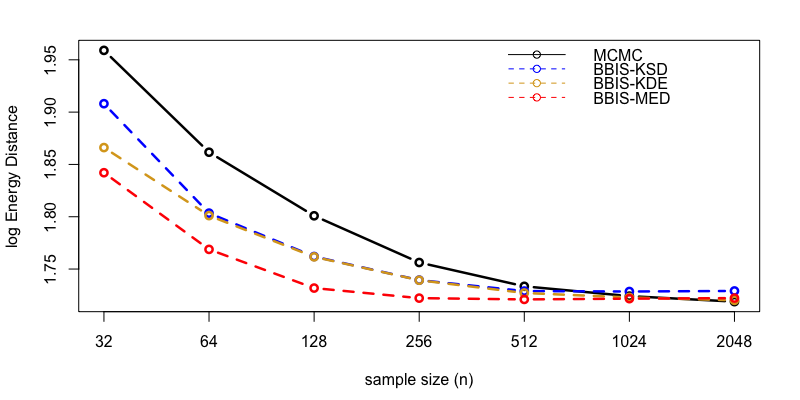}
    \caption{Energy Distance.}
\end{subfigure}
\caption{Comparison of the BBIS algorithms to the MCMC baseline on the $p=2$ dimensional mixture Gaussian example. Left panel shows its contour with the red dots indicating the 20 mixture centers. Right panel shows the energy distance \eqref{eq:mced} averaged over 100 independent runs.}
\label{fig:mixture_2d}
\end{figure}

\begin{figure}[!t]
\centering
\begin{subfigure}{0.48\textwidth}
    \centering
    \includegraphics[width=\textwidth]{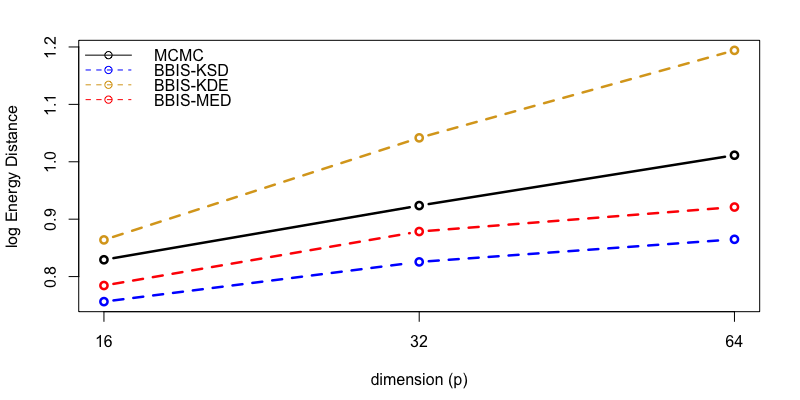}
    \caption{Energy Distance.}
\end{subfigure}
\begin{subfigure}{0.48\textwidth}
    \centering
    \includegraphics[width=\textwidth]{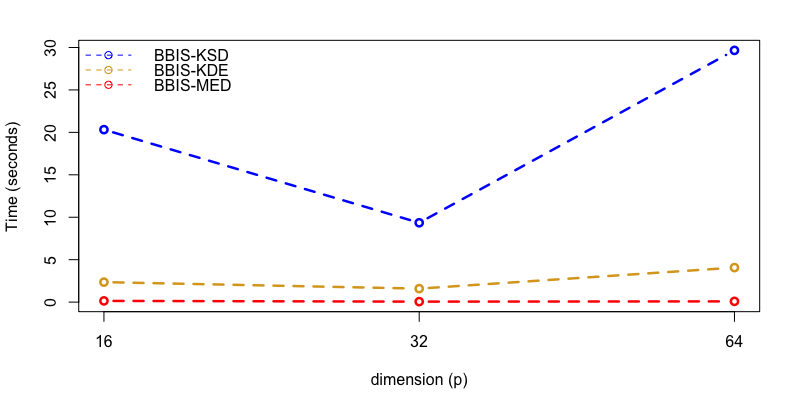}
    \caption{Computational Time.}
\end{subfigure}
\caption{Comparison of the BBIS algorithms to the MCMC baseline on the mixture Gaussian example with different dimension $p$. The energy distance \eqref{eq:mced} is averaged over 100 independent runs and normalized by dimension, i.e., divided by $\sqrt{p}$. The computational time is also averaged over 100 independent runs.}
\label{fig:mixture_hd}
\end{figure}

Now consider a complex multi-modal target distribution $\pi(x) = K^{-1}\sum_{k=1}^{K}\mathcal{N}(x;0,I_{p})$, the mixture Gaussian distribution. We use the same evaluation metric (energy distance), simulation mechanism (adaptive MCMC), and baseline (MCMC baseline) from the multivariate Gaussian example. Figure~\ref{fig:mixture_2d} shows the performance of the BBIS algorithms on a $p=2$ dimensional $K=20$ mixture example with centers randomly sampled from $\mathcal{N}(0,3^2 I_{2})$. We can see that BBIS-MED outperforms its competitors. Figure~\ref{fig:mixture_hd} shows the comparison on $p=16,32,64$ dimensional mixture Gaussian with centers randomly sampled from $\mathcal{N}(0,I_{p})$. Here $n=512$ MCMC samples are used. Again from left panel of Figure~\ref{fig:mixture_hd} we can observe BBIS-MED improves over the MCMC baseline but BBIS-KSD stands out as the winner in this higher dimensional problem. On the other hand from the right panel of Figure~\ref{fig:mixture_hd}, BBIS-MED only takes a fraction of time by BBIS-KSD. The marginal reduction in performance of BBIS-MED can be justified by its substantial computational saving.  

\subsection{Logistic Regression Example}
\label{subsec:logistic_regression_example}

\begin{figure}[!t]
\centering
\begin{subfigure}{0.48\textwidth}
    \centering
    \includegraphics[width=\textwidth]{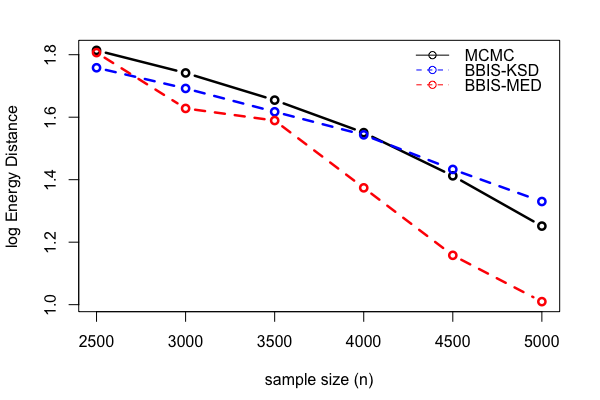}
    \caption{Energy Distance.}
\end{subfigure}
\begin{subfigure}{0.48\textwidth}
    \centering
    \includegraphics[width=\textwidth]{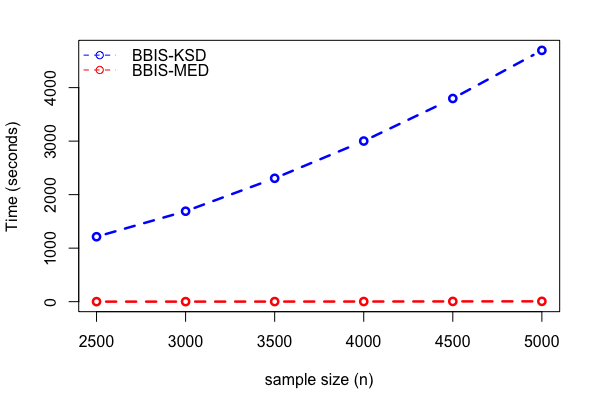}
    \caption{Computational Time.}
\end{subfigure}
\caption{Comparison of BBIS-MED and BBIS-KSD on the Logistic regression example. The energy distance \eqref{eq:mced} is computed with respect to the ``groundtruth" posterior samples.}
\label{fig:logistic}
\end{figure}

Consider the Bayesian logistic regression model for binary classification. Let $D=\{x_{i},y_{i}\}_{i=1}^{N}$ be the set of observed data with feature $x_{i}\in\mathbb{R}^{p}$ and label $y_{i}\in\{-1,1\}$. Let $\beta$ denotes the model coefficients. The log-posterior distribution is 
\begin{align*}
\log p(\beta|D) = \text{const.} + \log p(D|\beta) + \log p(\beta),
\end{align*}
where 
\begin{align*}
    \log p(D|\beta) = -\sum_{i=1}^{N}\log\left\{1 + \exp(-y_{i} x_{i}^{T}\beta)\right\},
\end{align*}
and $p(\beta) = \mathcal{N}(\beta;0,0.1\cdot I_{p})$ be the prior. We use the Forest Covtype dataset studied in \citep{liu2017bbis} and randomly select 50{,}000 data points for fitting the logistic regression. The dataset has 54 features and we also include an intercept term in the model, so this is a 55 dimensional problem. Given that we cannot analytically derive the posterior, we run the adaptive MCMC \citep{vihola2012mcmc} for 50{,}000 iterations and keep the second half 25{,}000 samples as the ``groundtruth" posterior samples. Figure~\ref{fig:logistic} shows the comparison of $n$ adaptive MCMC \emph{proposal} samples with post BBIS-MED and BBIS-KSD correction. Again the \emph{accepted} MCMC samples serve as the baseline for the comparison. We can see that BBIS-MED not only improves over the MCMC baseline for different $n$ in terms of the energy distance with respect to the ``groundtruth" posterior samples, but also beats the BBIS-KSD for larger $n$. Moreover, the significant computational saving of BBIS-MED is also observed: 6 seconds for BBIS-MED vs. 4{,}700 seconds for BBIS-KSD to compute the weight on $n=5{,}000$ samples, demonstrating that BBIS-MED is the efficient choice when the sample size is large. We do not include the BBIS-KDE in this study because of its poor performance compared to both BBIS-MED and BBIS-KSD in the previous two examples. 

\section{Bayesian Model Calibration}
\label{sec:bayesian_model_calibration}

\begin{figure}[!t]
\centering
\includegraphics[width=0.9\textwidth]{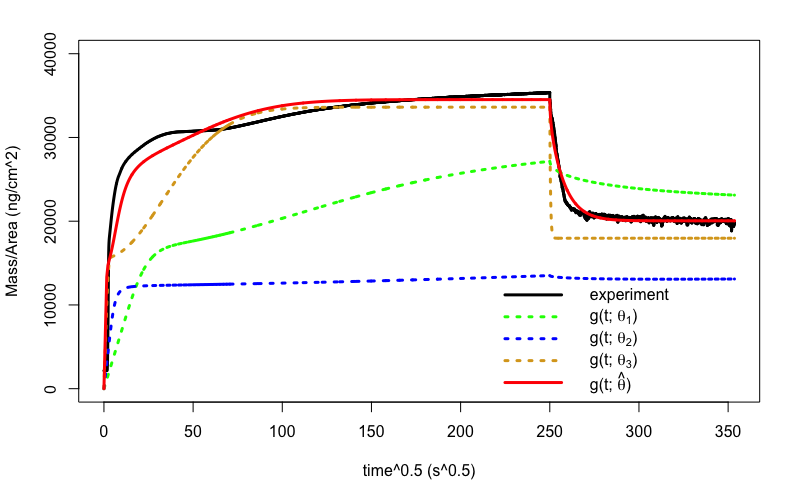}
\caption{The evaluation of the PDEs model $g$ on three randomly chosen $\theta$'s (dashed lines) versus the experimental observation (black solid line). The red solid line is the PDEs' output at $\hat{\theta}$, the posterior mean of the weighted samples by BBIS-MED.}
\label{fig:vpi_pde}
\end{figure}

Now we demonstrate the application of our proposed BBIS-MED algorithm on a real world model calibration problem where the score function of the posterior is difficult to compute. Vapor phase infiltration (VPI) process is one important class of chemical processes that can be used for transforming organic polymer membranes into hybrid organic-inorganic membranes \citep{leng2017vapor}. These hybrid membranes  enjoys 10 times energy efficiency than distillation for the chemical separation process, e.g. purification of clean water. Understanding the chemical behavior of the VPI process is critical for designing those membranes. \citet{ren2021vpi} proposed a reaction-diffusion partial differential equations (PDEs) model $g(t;\theta)$ for studying VPI, where $t$ denotes the time in seconds and $\theta$ denotes the model parameters. See Appendix \ref{appendix:vpi_model_calibration} for the model details. Figure~\ref{fig:vpi_pde} shows the evaluation of the PDEs model $g$ on some randomly chosen $\theta$'s, and none is aligned well with the experimental observation. Given that only temporal observations are available from experiment, i.e., spatial information is aggregated, this limits the choice of methods we can use to calibrate $\theta$. 

\begin{figure}[!t]
\centering
\includegraphics[width=0.9\textwidth]{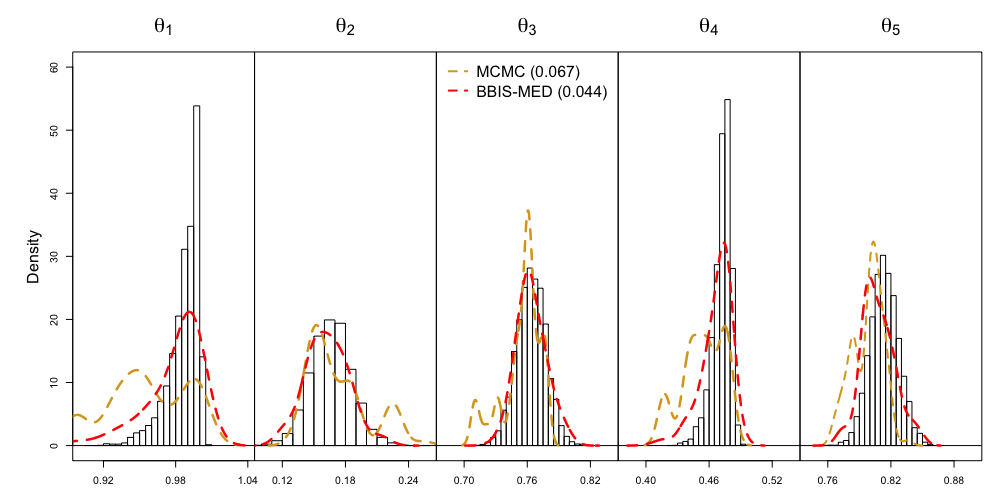}
\caption{Histogram of the ``groundtruth" posterior samples versus the densities (dashed lines) of the baseline MCMC samples and the BBIS-MED weighted samples. Number in the parentheses is the energy distance \eqref{eq:mced} to the ``groundtruth" posterior samples.}
\label{fig:vpi_posterior}
\end{figure}

One approach is to apply the Bayesian calibration \citep{kennedy2001calibration}. For simplicity assume there is no model discrepancy, thus the measurement errors are independent, and we can model them using i.i.d. Gaussian with mean zero and variance $\sigma^2$, i.e., the likelihood for $y(t)$, the noisy observation at time $t$, is
\begin{align*}
  y(t) \overset{i.i.d.}{\sim} \mathcal{N}(g(t;\theta), \sigma^2). 
\end{align*}
For the prior, we consider $p(\sigma^2) \propto 1/\sigma^2$ and $p(\theta)$ for the PDEs parameter (see Appendix \ref{appendix:vpi_model_calibration}). It follows that the posterior is 
\begin{align*}
  p(\theta,\sigma^2|y) \propto \frac{1}{\sigma^{N}}\exp\left\{-\frac{1}{2\sigma^2}\sum_{i=1}^{N}(y(t_{i}) - g(t_{i};\theta))^2\right\} \times p(\theta)p(\sigma^2), 
\end{align*}
where $\mathcal{T} = \{t_{1},\ldots,t_{N}\}$ are the 251 equally spaced out time points on $[0,125000]$ that we have both experimental observation and the PDEs model output. After integrating out $\sigma^2$ we have
\begin{align*}
  \log p(\theta|y) = \text{const.} - \frac{N}{2}\log\left\{\sum_{i=1}^{N}(y(t_{i}) - g(t_{i};\theta))^2\right\} + \log p(\theta). 
\end{align*}
Note that the PDEs model $g$ is in the log posterior, and hence the score function of $p(\theta|y)$ cannot be easily computed\footnote{Though for this particular PDEs example the score function can be computed via the adjoint method, it would still be computationally expensive. Moreover the purpose here is to demonstrate BBIS-MED can be easily applied without knowing the score function, which is common in many real world model calibration problems that the computer code is completely black-box.}. Moreover, PDEs is expensive to solve, and each evaluation of the log posterior takes about 7 seconds on a 2 cores 2.3 GHz laptop using the \texttt{deSolve} package in R \citep{soetaert2010desolve}. Given the limited computational resource, we can only afford to run the adaptive MCMC \citep{vihola2012mcmc} for 2{,}500 iterations, but this is far from converging by comparing the densities of the post burn-in MCMC samples (first half of the chain is discarded for burn-in) to the histograms of the ``groundtruth" posterior samples in Figure~\ref{fig:vpi_posterior}. The ``groundtruth" posterior samples are obtained by running adaptive MCMC for 25{,}000 iterations with the chain initialized at the mean of the post burn-in MCMC samples and the initial kernel covariance is set to be the diagonal of the covariance of the post burn-in MCMC samples. Again the first 5{,}000 samples are discarded for burn-in and the rest 20{,}000 samples are used for the ``groundtruth" posterior samples.

Fortunately, our proposed BBIS-MED algorithm could be the remedy. Applying BBIS-MED on the 2{,}500 MCMC \emph{proposal} samples we obtain the weighted samples that provide a much better approximation for the posterior (red dashed lines versus the histogram's in Figure~\ref{fig:vpi_posterior}). The improvement over the post burn-in MCMC samples can also be quantified by the energy distance (0.044 versus 0.067). Lastly, Figure~\ref{fig:vpi_pde} shows the PDEs output evaluated at the mean of the BBIS-MED weighted samples in red solid line, and it better matches to the experimental observations than the randomly chosen $\theta$'s.

\section{Conclusion}
\label{sec:conclusion}

In this paper we propose a novel Black-Box Importance Sampling algorithm, BBIS-MED, that is built upon the energy criterion \citep{joseph2015mined,joseph2019mined} motivated from the optimal configuration of charged particles in physics. Different from the kernelized Stein discrepancy based BBIS algorithm \citep{liu2017bbis}, our proposed BBIS-MED only require the unnormalized (log) density function, making it applicable to almost all Bayesian problems. Moreover, BBIS-MED also outperforms other BBIS algorithms that do not need the score function, e.g. the KDE-based BBIS algorithm \citep{delyon2016bbis}, and the unweighted baseline on extensive simulations. The order of magnitude reduction in computational time is yet another great benefit of our proposed BBIS-MED algorithm. 

Though the theoretical guarantee of BBIS-MED requires further study, we observe its outstanding empirical performances on a wide range of problems, including strongly correlated distributions (Figure~\ref{fig:gaussian_edist_tau}), high dimensional multi-modal distributions (Figure~\ref{fig:mixture_hd}), and the complicated posterior from real world applications (Figure~\ref{fig:logistic} and Figure~\ref{fig:vpi_posterior}). Moreover, the comparison of BBIS-MED to the \emph{Monte Carlo} baseline (Figure~\ref{fig:gaussian_mc_edist}) also further strengthen the validity of our proposed method. %We hope to provide some theoretical guidance in this manuscript, but the underlying mathematical problem is very challenging: look at the closely related problem of finding the best packing of spheres, which remained open for almost four centuries, but anyone who has ever stacked oranges already knew the answer \citep{borodachov2019energy}!

Although this paper focus on applying BBIS-MED to correct the bias from the finite MCMC samples, i.e., a simple yet efficient post-processing step to improve the approximation quality of any MCMC samples, BBIS-MED has much broader potential applications, including optimal MCMC thinning \citep{chopin2021thinning,riabiz2022thinning}, result refinement of complex variational proposals \citep{wainwright2008vi,rezende2015vi}, covariance shift in model training \citep{sugiyama2007cs}, extension to black-box importance weights for discrete variables (Appendix~\ref{appendix:discrete_variables}), and many other statistical and machine learning problems. 

\vspace{.25in}
\noindent {\Large\bf Acknowledgments}

\noindent This research is supported by  a U.S. National Science Foundation grant DMS-2310637.

\medskip

\bibliography{references}

\newpage

\setcounter{table}{0}
\renewcommand{\thetable}{S\arabic{table}}%
\setcounter{figure}{0}
\renewcommand{\thefigure}{S\arabic{figure}}%
\setcounter{section}{0}

\section*{\LARGE{Appendices}}

\appendix

\section[Additional Simulation Results for Multivariate Gaussian Example]{Additional Simulation Results for the \\ Multivariate Gaussian Example}
\label{appendix:additional_simulation_multivariate_gaussian}

We present the performance of BBIS-MED on some other evaluation metrics for the multivariate Gaussian example in Subsection~\ref{subsec:multivariate_gaussian_example}. We consider the approximation error (mean squared error) on the following test functions $\mathbb{E}_{\pi}X$, $\mathbb{E}_{\pi}\text{Diag}(XX^{T})$, and $\mathbb{E}_{\pi}\sin(1^{T}X)$ where $X\sim\pi=\mathcal{N}(0,\Sigma)$. The mean squared error for the first two metrics are averaged over the components. Similar results are observed: BBIS-MED are comparable to BBIS-KSD and generally outperforms the MCMC baseline and BBIS-KDE. 

\begin{figure}[b!]
\centering
\begin{subfigure}{0.32\textwidth}
    \centering
    \includegraphics[width=\textwidth]{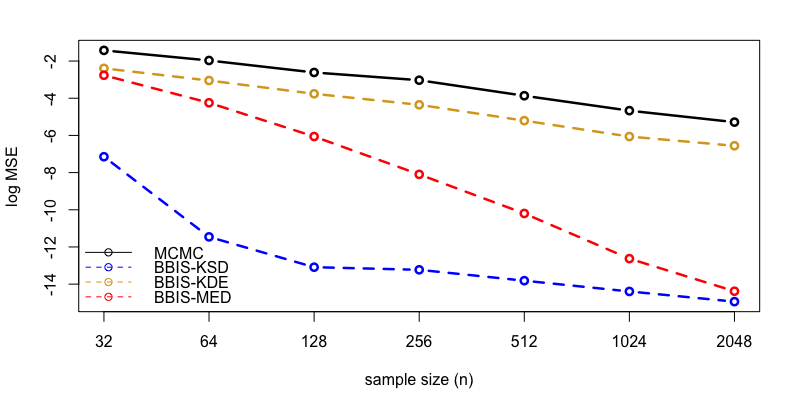}
    \caption{$\mathbb{E}_{\pi} X$}
\end{subfigure}
\begin{subfigure}{0.32\textwidth}
    \centering
    \includegraphics[width=\textwidth]{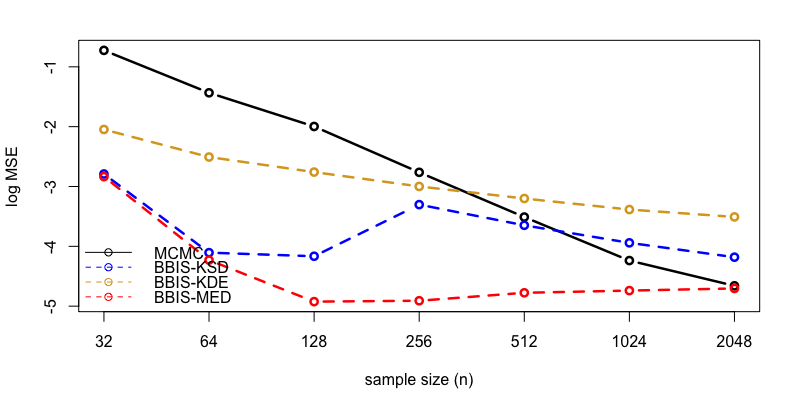}
    \caption{$\mathbb{E}_{\pi} \text{Diag}(XX^{T})$}
\end{subfigure}
\begin{subfigure}{0.32\textwidth}
    \centering
    \includegraphics[width=\textwidth]{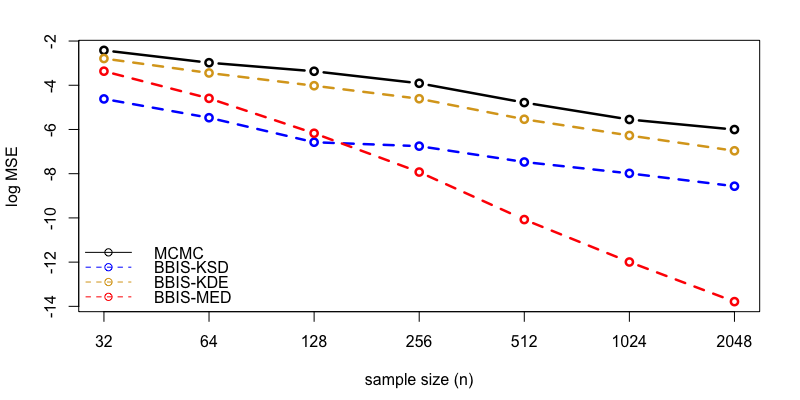}
    \caption{$\mathbb{E}_{\pi} \sin(1^{T}X)$}
\end{subfigure}
\caption{Comparison of the BBIS algorithms to the MCMC baseline on the multivariate Gaussian example with different sample size $n$. Dimension $p=2$ and correlation parameter $\tau=0$. Mean square error (MSE) is computed averaged over 100 runs.}
\end{figure}

\begin{figure}[b!]
\centering
\begin{subfigure}{0.32\textwidth}
    \centering
    \includegraphics[width=\textwidth]{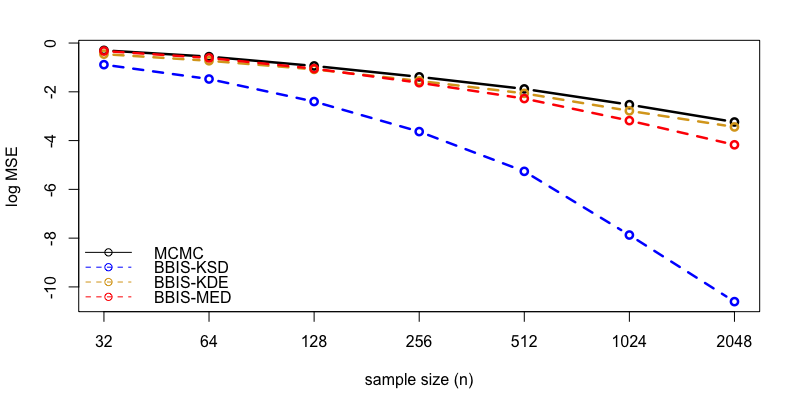}
    \caption{$\mathbb{E}_{\pi} X$}
\end{subfigure}
\begin{subfigure}{0.32\textwidth}
    \centering
    \includegraphics[width=\textwidth]{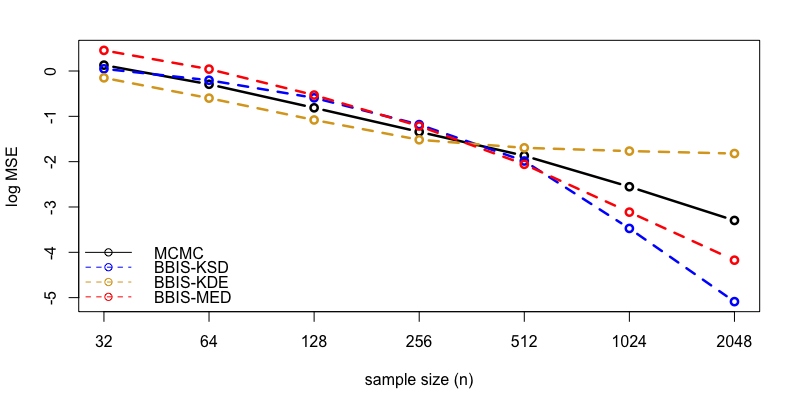}
    \caption{$\mathbb{E}_{\pi} \text{Diag}(XX^{T})$}
\end{subfigure}
\begin{subfigure}{0.32\textwidth}
    \centering
    \includegraphics[width=\textwidth]{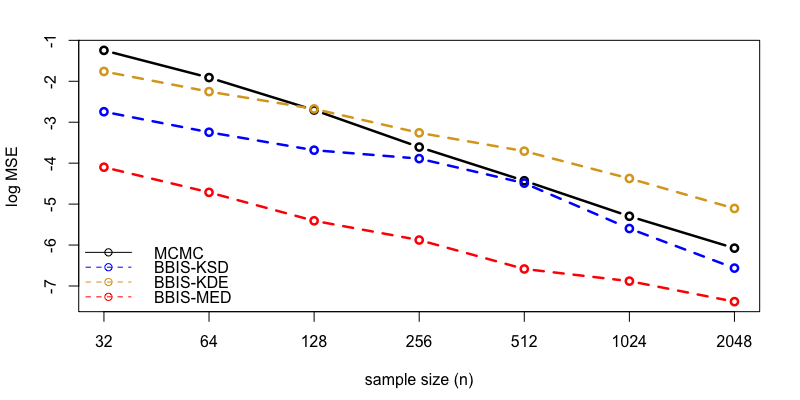}
    \caption{$\mathbb{E}_{\pi} \sin(1^{T}X)$}
\end{subfigure}
\caption{Comparison of the BBIS algorithms to the MCMC baseline on the multivariate Gaussian example with different sample size $n$. Dimension $p=16$ and correlation parameter $\tau=0$. Mean square error (MSE) is computed averaged over 100 runs.}
\end{figure}

\begin{figure}[t!]
\centering
\begin{subfigure}{0.32\textwidth}
    \centering
    \includegraphics[width=\textwidth]{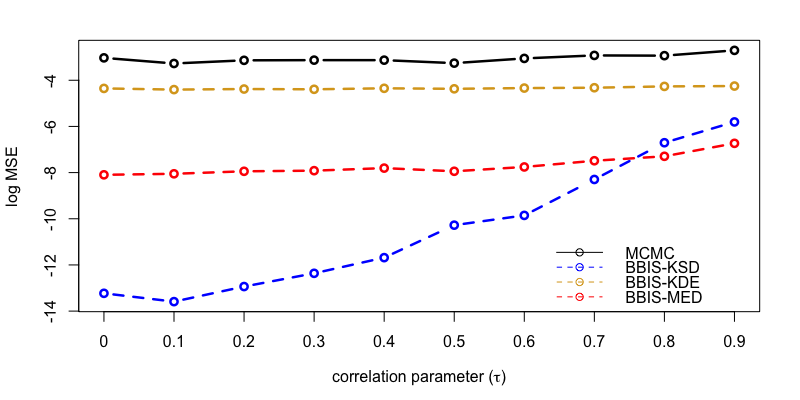}
    \caption{$\mathbb{E}_{\pi} X$}
\end{subfigure}
\begin{subfigure}{0.32\textwidth}
    \centering
    \includegraphics[width=\textwidth]{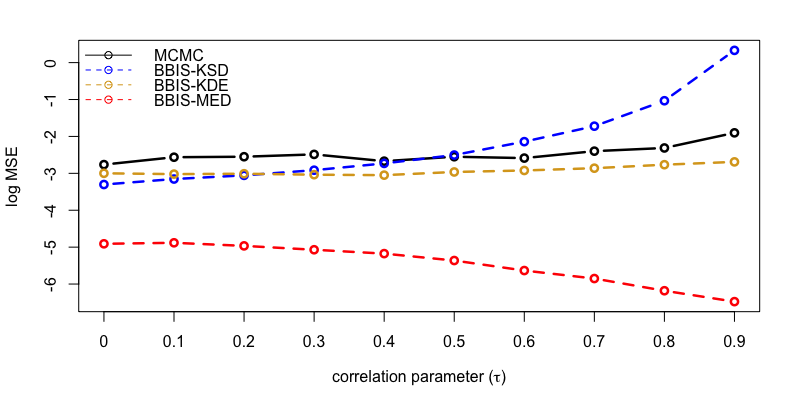}
    \caption{$\mathbb{E}_{\pi} \text{Diag}(XX^{T})$}
\end{subfigure}
\begin{subfigure}{0.32\textwidth}
    \centering
    \includegraphics[width=\textwidth]{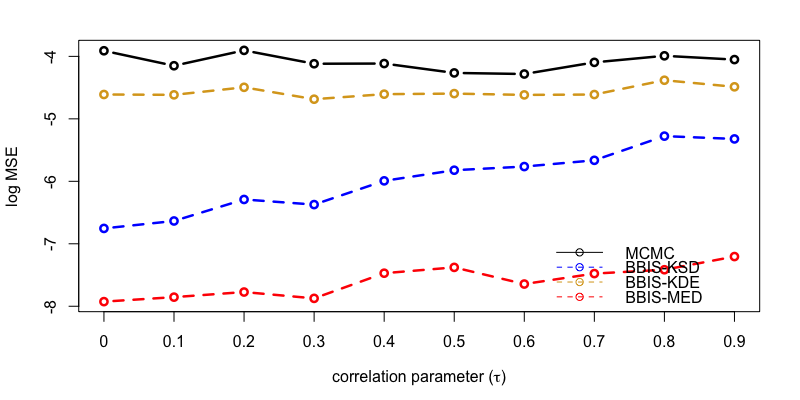}
    \caption{$\mathbb{E}_{\pi} \sin(1^{T}X)$}
\end{subfigure}
\caption{Comparison of the BBIS algorithms to the MCMC baseline on the multivariate Gaussian example with different correlation parameter $\tau$. Dimension $p=2$ and sample size $n=256$. Mean square error (MSE) is computed averaged over 100 runs.}
\end{figure}

\begin{figure}[t!]
\centering
\begin{subfigure}{0.32\textwidth}
    \centering
    \includegraphics[width=\textwidth]{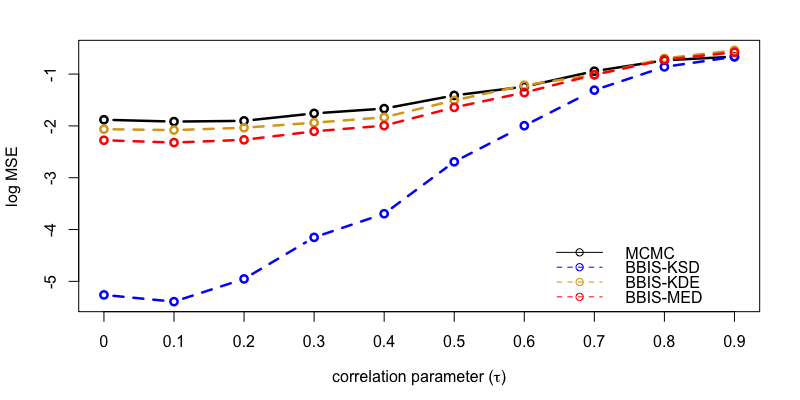}
    \caption{$\mathbb{E}_{\pi} X$}
\end{subfigure}
\begin{subfigure}{0.32\textwidth}
    \centering
    \includegraphics[width=\textwidth]{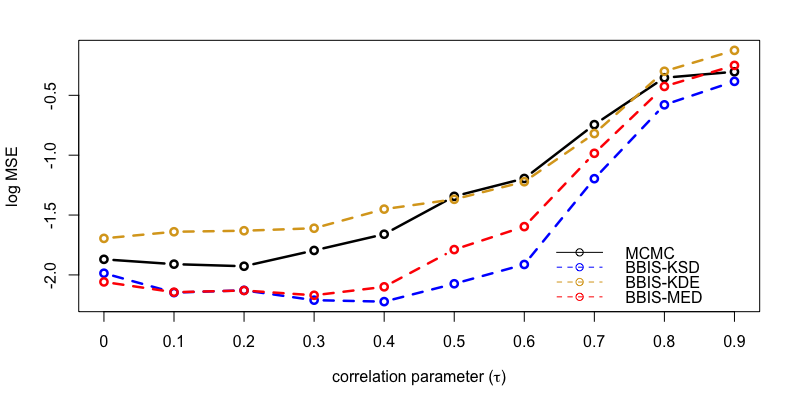}
    \caption{$\mathbb{E}_{\pi} \text{Diag}(XX^{T})$}
\end{subfigure}
\begin{subfigure}{0.32\textwidth}
    \centering
    \includegraphics[width=\textwidth]{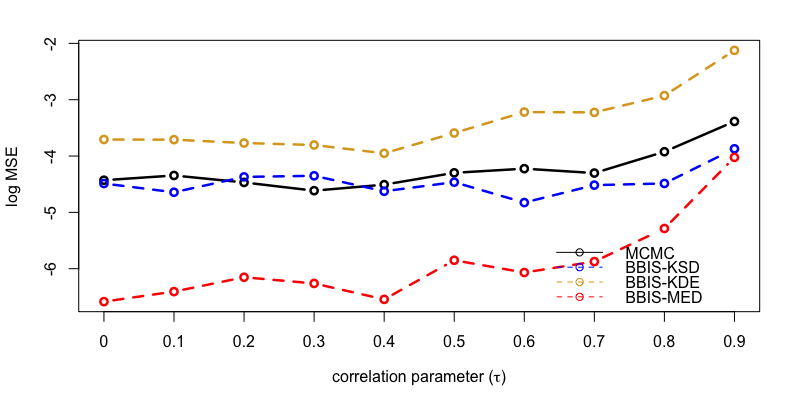}
    \caption{$\mathbb{E}_{\pi} \sin(1^{T}X)$}
\end{subfigure}
\caption{Comparison of the BBIS algorithms to the MCMC baseline on the multivariate Gaussian example with different correlation parameter $\tau$. Dimension $p=16$ and sample size $n=512$. Mean square error (MSE) is computed averaged over 100 runs.}
\end{figure}

\begin{figure}[t!]
\centering
\begin{subfigure}{0.32\textwidth}
    \centering
    \includegraphics[width=\textwidth]{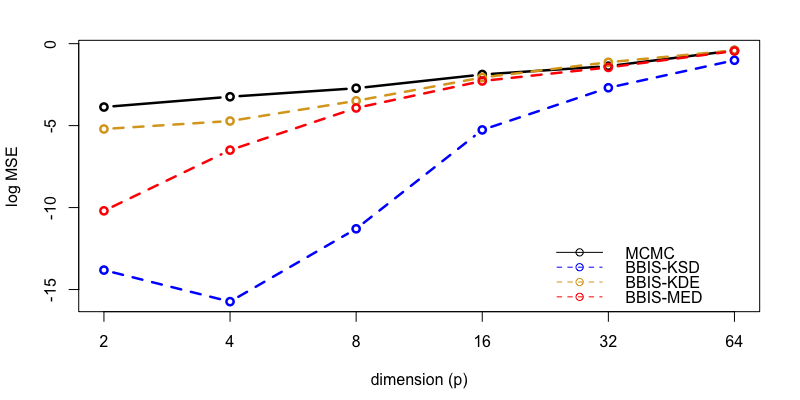}
    \caption{$\mathbb{E}_{\pi} X$}
\end{subfigure}
\begin{subfigure}{0.32\textwidth}
    \centering
    \includegraphics[width=\textwidth]{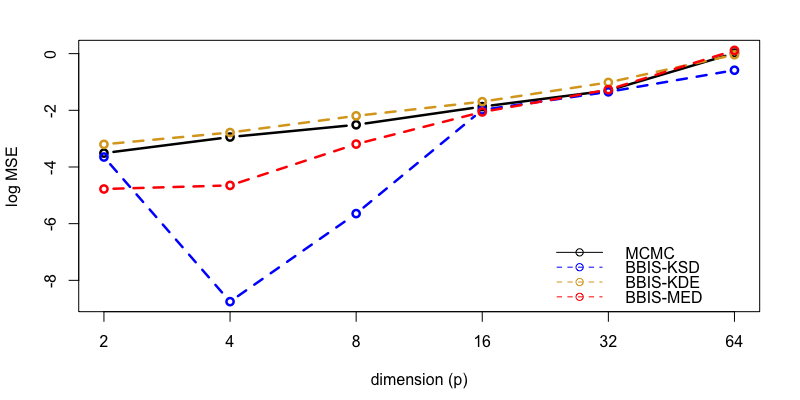}
    \caption{$\mathbb{E}_{\pi} \text{Diag}(XX^{T})$}
\end{subfigure}
\begin{subfigure}{0.32\textwidth}
    \centering
    \includegraphics[width=\textwidth]{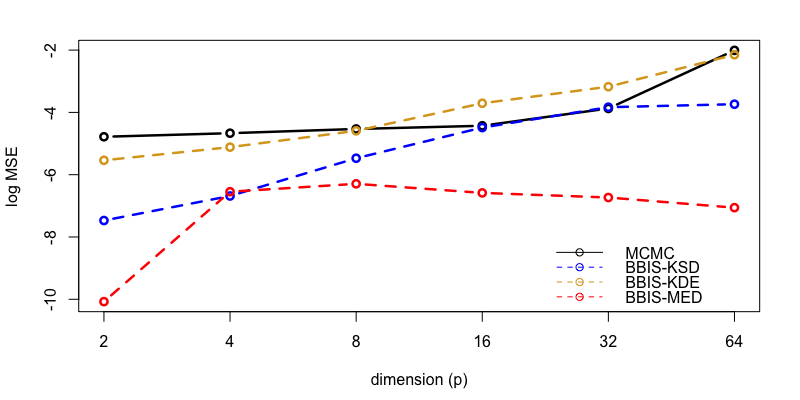}
    \caption{$\mathbb{E}_{\pi} \sin(1^{T}X)$}
\end{subfigure}
\caption{Comparison of the BBIS algorithms to the MCMC baseline on the multivariate Gaussian example with different dimension $p$. Sample size $n=512$ and correlation parameter $\tau=0$. Mean square error (MSE) is computed averaged over 100 runs.}
\end{figure}

\section{Implementation Details of BBIS-MED}
\label{appendix:implementation_details}

Recall that BBIS-MED finds the optimal importance weight $w^{*}$ by solving  
\begin{align}
  \label{aeq:meis}
  w^{*} = \arg\min_{w\in\Delta^{n}} w^{T}Rw,
\end{align}
where 
\begin{align}
  \label{aeq:meis_ij}
  R_{ij} = \exp\left\{-k\left(\frac{1}{2p}\log\pi(x_{i})+\frac{1}{2p}\log\pi(x_{j}) + \frac{1}{2}\log(d^{2}(x_i,x_j)+\delta)\right)\right\}.
\end{align} 

\subsection[Positive Definite of R]{Positive Definiteness of $R$}
\label{appendix:pd_of_R}
Let us show that the matrix $R$ in \eqref{aeq:meis} is positive definite. Let $K$ denotes the matrix by evaluating the inverse multiquadric kernel on samples $\{x_{i}\}_{i=1}^{n}$, i.e., $K_{ij} = (d(x_i,x_j)^2+\delta)^{-1/2}$ for some $\delta>0$, and $P = \text{Diag}(\rho(x_{1}),\ldots,\rho(x_{n}))\neq \mathbf{0}$, we have 
\begin{align*}
  R = \tilde{R}^{(k)}, \quad \tilde{R} = P^{T}KP,
\end{align*}
where $\tilde{R}^{(k)} = (\tilde{R}_{ij}^{k})$ is the Hadamard power. It is also called the elementwise/Schur power. Given that $K$ is the positive definite, then for all $a\in\mathbb{R}^{n}\backslash\{0\}$, we have $a^{T}Ka > 0$. It follows that for any $b\in\mathbb{R}^{n}\backslash\{0\}$, we have
\begin{align*}
  b^{T}\tilde{R} b = b^{T}P^{T}KPb = (Pb)^{T}K(Pb) = a^{T}Ka > 0,
\end{align*}
where $a = Pb\in\mathbb{R}^{n}\backslash\{0\}$. By the Schur Product Theorem \citep{schur1911psd}, $R=\tilde{R}^{(k)}$ is positive definite for any positive integer $k$. Moreover, we have the following result from \citet{horn1969psd} and \citet{fitzgerald1977psd}: if $f:(0,\infty)\to\mathbb{R}$ is a smooth function such that the matrix $(f(a_{ij}))$ is positive definite whenever $A=(a_{ij})$ is a $n\times n$ positive definite matrix with positive entries and $f,f^{\prime},f^{\prime\prime},\ldots,f^{(n-1)}$ are all nonnegative on $\mathbb{R}^{+}$. Here we have (i) $\tilde{R}_{ij} > 0$, and (ii) $f(x) = x^{k}$ is smooth and $f,f^{\prime},f^{\prime\prime},\ldots,f^{(n-1)}$ are all nonnegative on $\mathbb{R}^{+}$ if $k>(n-2)$. Hence it follows that $R=\tilde{R}^{(k)}$ is also positive definite for any positive real number $k>(n-2)$. It follows that the optimization in \eqref{aeq:meis} is a convex quadratic programming problem if $k$ is a positive integer or any real number greater than $n-2$. 

\subsection[Choice for k]{Choice for $k$}
\label{appendix:choice_for_k}

\begin{figure}[t!]
\centering
\begin{subfigure}{0.32\textwidth}
    \centering
    \includegraphics[width=\textwidth]{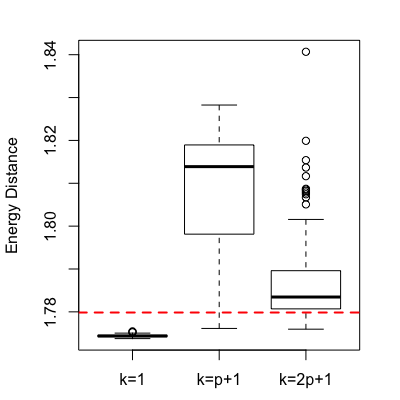}
    \caption{$p=2$}
\end{subfigure}
\begin{subfigure}{0.32\textwidth}
    \centering
    \includegraphics[width=\textwidth]{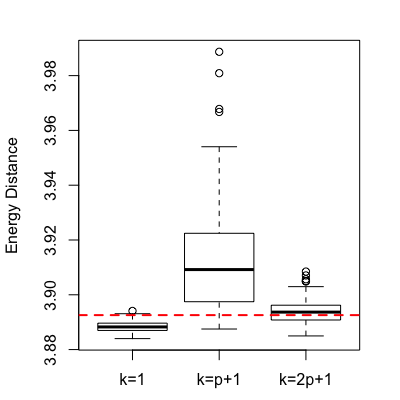}
    \caption{$p=8$}
\end{subfigure}
\begin{subfigure}{0.32\textwidth}
    \centering
    \includegraphics[width=\textwidth]{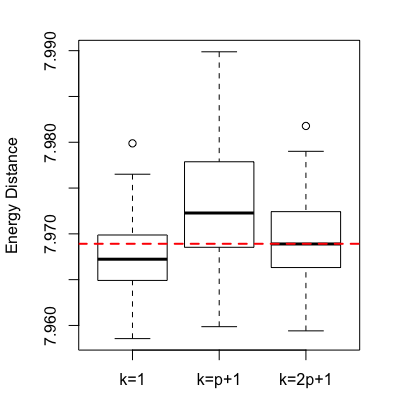}
    \caption{$p=32$}
\end{subfigure}
\caption{Comparison of BBIS-MED \eqref{eq:meis} with $k=1, p+1, 2p+1$ on the multivariate Gaussian example with $n=256$ and $\tau=0$. The boxplots summarize the energy distance \eqref{eq:mced} over 100 independent runs. The red dashed lines are the average energy distance \eqref{eq:mced} of the MC baseline (unweighted samples simulated directly from multivariate Gaussian) over 100 independent runs.}
\label{fig:choice_k}
\end{figure}

We focus on the case where $k$ is a positive integer such that the matrix $R$ in \eqref{aeq:meis} is positive definite. Based on the theoretical result of MED \citep{borodachov2008riesz1,borodachov2008riesz2,wang2011med,joseph2015mined}, one should consider $k>p$ to have the limiting distribution converging to the target distribution. However, empirically from Figure~\ref{fig:choice_k} we see that $k=1$ yields better performance than both $k=p+1$ and $k=2p+1$ on different $p$'s. We conjecture this could be due to the inclusion of the self-interaction terms in the energy criterion computation. Though existing empirical results suggest that $k=1$ is a good robust choice, further theoretical investigation is needed to understand the optimal choice for $k$.

\subsection{Improving Numerical Robustness}
\label{appendix:improving_numerical_robustness}

By using the projected gradient descent \citep{duchi2008projgrad} to solve \eqref{aeq:meis}, we notice that numerical instability could occur if some of the samples have very small density value, which leads to a few very large terms in the diagonal of $R$ in \eqref{aeq:meis}. One simple solution is to first remove those very low-density samples and assigning them zero weights, since those samples are not important for the approximation and would be assigned weight close to zero in the optimum solution. The problem is to identify the correct threshold $\nu$ for the cutoff, i.e., we only compute weights for $\{x_{i}:\log\pi(x_{i})\geq\nu\}$ and assign 0 weight for the rest samples. This also reduce the computational complexity of solving \eqref{aeq:meis}. After standardizing the data to have component-wise zero mean and unit variance, we consider 
\begin{align*}
  \nu = \max_{i=1,\ldots,n}\log\pi(x_{i}) - p((\delta)^{-1/2} - ((20\sqrt{p})^2+\delta)^{-1/2}) - 1,
\end{align*}
that works quite well in the simulation studies when $k=1$. 

\section{VPI Process Model Calibration}
\label{appendix:vpi_model_calibration}

To study the VPI process, \citet{ren2021vpi} propose the following partial differential equations (PDEs), 
\begin{align*}
  \left\{\begin{array}{l}
    \frac{\partial C_{free}}{\partial t} = D\frac{\partial^2 C_{free}}{\partial x^2} - kC_{free}C_{polymer} \\
    \frac{\partial C_{product}}{\partial t} = kC_{free}C_{polymer} \\
    D = D_0 \exp(-K^{\prime}C_{product}) \\
    \frac{\partial C_{polymer}}{\partial t} = -kC_{free}C_{polymer} \\
  \end{array}\right. 
\end{align*}
and the initial and boundary conditions are
\begin{align*}
  \left\{\begin{array}{lll}
    C_{free} = 0, & 0 < x < l, & t = 0 \\
    C_{product} = 0, & 0 < x < l, & t = 0 \\
    C_{polymer} = C_{polymer}^{0}, & 0 < x < l, & t = 0 \\
    \frac{\partial C_{free}}{\partial x} = 0, & x = 0, & t > 0 \\
    C_{free} = C_{s}, & x = l, & t > 0 \\
  \end{array}\right.
\end{align*}
where $C_{free} (\textrm{mol}/\textrm{cm}^3)$ is the concentration of the free diffusing vapor-phase precursor, $C_{polymer} (\textrm{mol}/\textrm{cm}^3)$ is the concentration of the accessible reactive polymeric functional groups, and $C_{product} (\textrm{mol}/\textrm{cm}^3)$ is the concentration of the immobilized product formed from the reaction between the free diffusing vapor-phase precursor and the polymeric functional groups. There are five unknown parameters $\theta = \{D_{0},C_{s},C_{polymer}^{0},K^{\prime},k\}$ in the PDEs: $D_{0} (\textrm{cm}^2/\textrm{s})$ is initial diffusivity of the free diffusing vapor-phase precursor, $C_{s} (\textrm{mol}/\textrm{cm}^3)$ is the surface concentration of the free diffusing vapor-phase precursor, $C_{polymer}^{0} (\textrm{mol}/\textrm{cm}^3)$ is the initial concentration of accessible reactive polymeric functional groups, $K^{\prime} (\textrm{cm}^3/\textrm{mol})$ is the hindering factor describing how the immobilized product slows down the diffusivity of free diffusing vapor, and $k (\textrm{cm}^3/\textrm{mol}\cdot\textrm{s})$ is the associated reaction rate. The parameter $l$ im the PDEs represents the polymer thickness, which is controllable in the experiment. For the experimental data that we use for the calibration (black solid line in Figure~\ref{fig:vpi_pde}), the polymer thickness $l = 483\textrm{nm}$.

Let $y(t)$ denotes the noisy in situ experimental measurements of the mass uptake (black solid line in Figure~\ref{fig:vpi_pde}) of the free diffusing vapor-phase precursor and the immobolized product from the chemical reaction over time, and $g(t;\theta)$ denotes the time series of $C_{free} + C_{product}$ from the PDEs' output (color dashed lines in Figure~\ref{fig:vpi_pde}). Assume that there is no model discrepancy, the calibration problem reduces to a nonlinear regression problem, 
\begin{align*}
  y(t) = g(t;\theta) + \epsilon_{t},
\end{align*}
where $\epsilon_{t}\overset{i.i.d.}{\sim}(0,\sigma^2)$ models the measurement error. Let the prior for $\sigma^2$ be $p(\sigma^2)\propto 1/\sigma^2$. For $\theta$, we consider the prior $p(\theta)=p(\log D_{0})p(D_{0})p(C_{s})p(C_{polymer}^{0})p(K^{\prime})p(\log k)$, and
\begin{align*}
  \log D_{0} &\sim \tilde{p}(\log D_{0}; \log(\textrm{1e-12}), \log(\textrm{1e-9}), 1000, 1000) \\
  C_{s} & \sim \tilde{p}(C_{s}; \textrm{4e-3}, \textrm{5e-3}, 1000, 1000) \\
  C_{polymer}^{0} & \sim \tilde{p}(C_{polymer}^{0}; \textrm{5e-3}, \textrm{6e-3}, 1000, 1000) \\
  K^{\prime} & \sim \tilde{p}(K^{\prime}; \textrm{500}, \textrm{2500}, 1000, 10) \\
  \log k & \sim \tilde{p}(\log k; \log(\textrm{1e-3}), \log(\textrm{1e1}), 1000, 1000)
\end{align*}
where 
\begin{align*}
  \tilde{p}(x;a,b,\lambda_{a},\lambda_{b}) = \exp\{\lambda_a(x-a)\}I(x<a) + I(a\leq x\leq b) + \exp\{-\lambda_b(x-b)\}I(x>b)
\end{align*}
is the prior distribution with the uniform prior for $x\in[a,b]$, the interval that we are certain about including the true parameter, and the exponential prior for $x\notin[a,b]$. We use the range suggested in \citep{huang2021bofo} that calibrate the VPI process by Bayesian optimization. For almost all parameters in the PDEs, we have good domain knowledge about its feasible range, and hence 1{,}000 is used for the exponential parameter in the prior. However, for the hindering factor $K^{\prime}$ that is less studied in the literature, we set the exponential parameter to be 10 for its upper bound in the prior. Last, we re-scale the parameters to be in the unit hypercube by their promising region $[a,b]$'s.

\section{Discrete Variables}
\label{appendix:discrete_variables}

Last, we want to discuss how we can adapt our proposed BBIS-MED algorithm for discrete (ordinal/nominal) variables, while such adaptation is not as straightforward for the competitor BBIS-KSD \citep{liu2017bbis}. The key idea lies in transforming the ordinal variable to continuous variable via scoring \citep{wu2011experiments}. For the nominal variable, we need to first convert them into numerical coding \citep{chambers1992helmert} for the distance computation. Similar approach is considered in \citep{joseph2022split} to compute the energy distance for nominal variable. 

\begin{figure}[!t]
\centering
\includegraphics[width=0.6\textwidth]{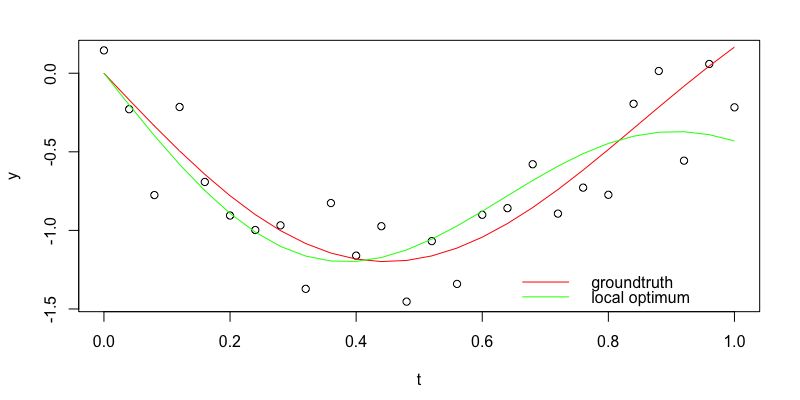}
\caption{The circle represents the noisy observation $y(t)$. The red solid line is the groundtruth function $h(t;x_{1}^{*},x_{2}^{*},\rho^{*})$ with $x_{1}^{*} = 0.719, x_{2}^{*}=0.552,\rho^{*}=3$. The green solid line is a local optimum solution $h(t;x_{1}^{\prime},x_{2}^{\prime},\rho^{\prime})$ with $x_{1}^{\prime}=0.997,x_{2}^{\prime}=0.625,\rho^{\prime}=2$.}
\label{fig:discrete_obs}
\end{figure}

Discrete variables are often observed in many Bayesian problems, including latent class mixture model and calibration of computer code. We focus again on the calibration application because the problem dimension is usually small or moderate, but the posterior evaluation is very expensive so only few hundreds/thousands MCMC samples are affordable. This is the setting where the improvement by BBIS algorithms are most significant from the empirical results. Let us consider the following functional-output computer code with two continuous input variables $x_{1},x_{2}\in[0,1]$, and one discrete variable $\rho\in\{0,1,2,3\}$:
\begin{align*}
    h(t;x_{1},x_{2},\rho) = x_{1}\sin\left(2\pi\frac{\rho}{7}t - \pi\right) + x_{2}\sin\left(2\pi\left(1-\frac{\rho}{7}\right)t - \pi\right),
\end{align*}
for $t\in[0,1]$. The observed data $y$ is generated by 
\begin{align*}
y(t) = h(t;x_{1}^{*},x_{2}^{*},t^{*}) + \epsilon(t)
\end{align*}
at $t=\{0.00,0.04,0.08,\ldots,0.96,1.00\}$, 26 evenly spaced points in $[0,1]$, and $\epsilon(t)\sim\mathcal{N}(0,0.2^2)$. We use $x_{1}^{*} = 0.719, x_{2}^{*}=0.552,\rho^{*}=3$ for this synthetic example. See Figure~\ref{fig:discrete_obs} for the synthetic observations we want to calibrate the parameter $x_{1},x_{2},t$ to. This is a difficult problem because there is a local optimum at $x_{1}^{\prime}=0.997,x_{2}^{\prime}=0.625,\rho^{\prime}=2$ (green line in Figure~\ref{fig:discrete_obs}).

\begin{figure}[t!]
\centering
\begin{subfigure}{0.32\textwidth}
    \centering
    \includegraphics[width=\textwidth]{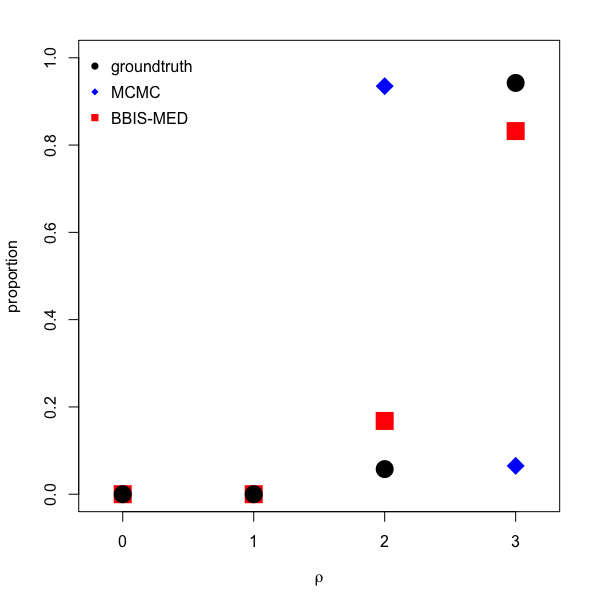}
    \caption{$\rho$}
\end{subfigure}
\begin{subfigure}{0.32\textwidth}
    \centering
    \includegraphics[width=\textwidth]{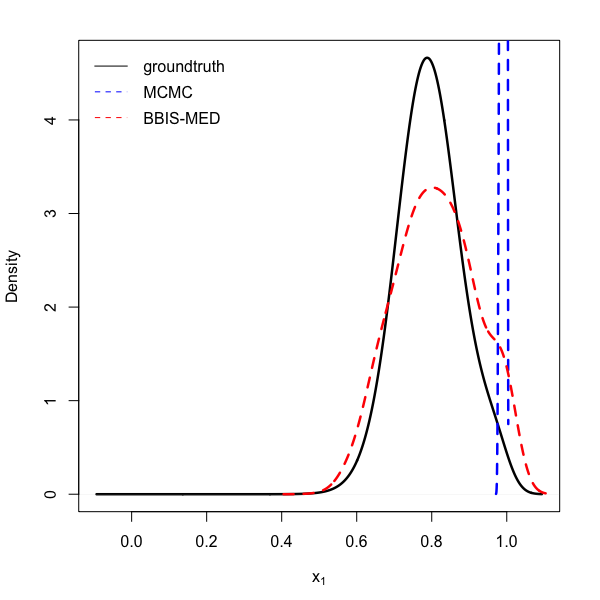}
    \caption{$x_{1}$}
\end{subfigure}
\begin{subfigure}{0.32\textwidth}
    \centering
    \includegraphics[width=\textwidth]{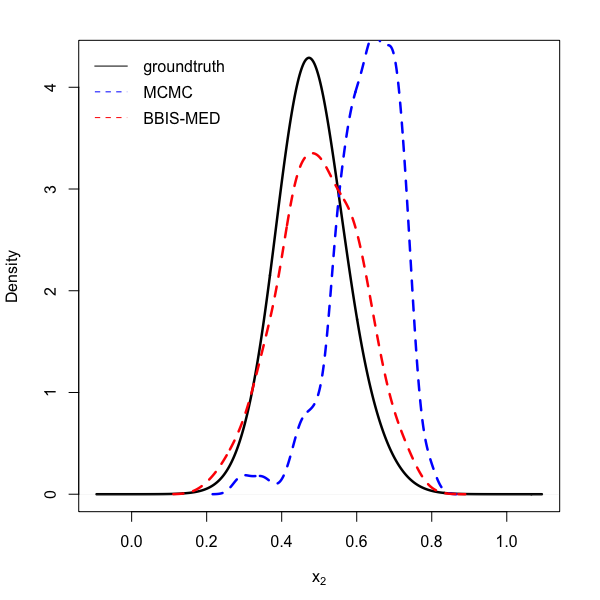}
    \caption{$x_{2}$}
\end{subfigure}
\caption{Groundtruth posterior (in black) vs. posterior estimated by MCMC (in blue) and BBIS-MED (in red) on the synthetic calibration problem with discrete variable $\rho$.}
\label{fig:discrete_posterior}
\end{figure}

This is similar to the model calibration problem we presented in Section~\ref{sec:bayesian_model_calibration}, so we can easily derive the log-posterior, i.e.,
\begin{align*}
    \log p(x_{1},x_{2},\rho|y) =& \text{const.} - \frac{N}{2}\log\left\{\sum_{i=1}^{N}(y(t_{i}) - h(t_{i};x_{1},x_{2},\rho))^2 \right\} + \\
    &\log p(x_{1}) + \log p(x_{2}) + \log p(\rho),
\end{align*}
where the $\mbox{Uniform}[0,1]$ prior is used for the continuous variable $x_{1},x_{2}$, and uniform discrete prior is used for $\rho$, i.e., $p(\rho=k)=1/4$ for all $k=0,1,2,3$. Given the number of parameters is small, we can obtain precise approximation of the posterior using importance sampling with prior $p(x_{1},x_{2},\rho)$ for the proposal distribution and the posterior $p(x_{1},x_{2},\rho|y)$ for the target distribution. We use 40{,}000 importance samples after normalization as the groundtruth posterior samples. Figure~\ref{fig:discrete_posterior} shows the groundtruth posterior in black solid line, and it matches closely with the $x_{1}^{*},x_{2}^{*},\rho^{*}$ that we use for this synthetic example. 

Now we describe the MCMC procedure. We are going to use Gibbs-type procedure: 
\begin{itemize}
    \item sample continuous variable $x_{1},x_{2}|y,\rho$ by Metropolis-Hasting with proposal $\mathcal{N}(0,0.1^2 I_{2})$.
    \item sample discrete variable $\rho|y,x_{1},x_{2}$ where the conditional distribution can be analytically computed, i.e., 
    \begin{align*}
        p(\rho=k|y,x_{1},x_{2}) = \frac{p(y,x_{1},x_{2},\rho=k)}{p(y,x_{1},x_{2})} = \frac{p(y,x_{1},x_{2},\rho=k)}{\sum_{i=0}^{3}p(y,x_{1},x_{2},\rho=i)}.
    \end{align*}
\end{itemize}
We run the MCMC for 2{,}000 iterations and discard the first half for burn-in. The marginal distribution of the 1{,}000 post burn-in MCMC samples is plotted in blue dashed line in Figure~\ref{fig:discrete_posterior}. We can see that the MCMC samples get trapped at the local optimum solution and cannot provide a accurate approximation for the posterior.

Given the discrete variable $\rho$ only has 4 levels, a better way to fully explore the posterior is to run adaptive MCMC \citep{vihola2012mcmc} on the continuous variable $x_{1},x_{2}$ for each $\rho$, and hence in total of 4 Markov chains are run. However, the stationary distribution of these Markov chains will be different from the target posterior distribution, but BBIS-MED can be used to correct for this distributional mismatch. We run the adaptive MCMC for 250 iterations for each $\rho$ and use all the proposal samples from the chains to compute the black-box weights. Total of 1{,}000 weighted samples are used for the posterior approximation, and from Figure~\ref{fig:discrete_posterior} we can see that BBIS-MED samples (in red dashed line) do not get trapped by the local optimum, and approximate the groundtruth posterior well.

\end{document}